\documentclass[12pt]{article}
\usepackage{graphics,epsfig}

\def\dalemb#1#2{{\vbox{\hrule height .#2pt
        \hbox{\vrule width.#2pt height#1pt \kern#1pt
                \vrule width.#2pt}
        \hrule height.#2pt}}}

\let\a=\alpha \let\b=\beta \let\g=\gamma \let\d=\delta \let\e=\epsilon
\let\z=\zeta  \let\th=\theta  \let\k=\kappa
\let\l=\lambda \let\m=\mu \let\n=\nu \let\x=\xi \let\p=\pi 
\let\s=\sigma \let\t=\tau   \let\c=\chi 
\let\vp=\varphi \let\vep=\varepsilon
\let\w=\omega      \let\G=\Gamma \let\D=\Delta \let\Th=\Theta \let\L=\Lambda
\let\X=\Xi \let\P=\Pi \let\S=\Sigma  \let\Y=\Psi
\let\C=\Chi \let\W=\Omega
\let\la=\label \let\ci=\cite 
  
\def\nn{\nonumber} \def\bd{\begin{document}} \def\ed{\end{document}}
\def\ds{\documentstyle} \let\fr=\frac \let\bl=\bigl \let\br=\bigr
\let\Br=\Bigr \let\Bl=\Bigl
\let\bm=\bibitem
\let\na=\nabla
\def\tU{{\widetilde U}}
\let\pa=\partial \let\ov=\overline
\def\ie{{\it i.e.\ }}
\newcommand{\be}{\begin{equation}}
\newcommand{\ee}{\end{equation}}
\def\ba{\begin{array}}
\def\ea{\end{array}}
\def\ft#1#2{{\textstyle{{\scriptstyle #1}\over {\scriptstyle #2}}}}
\def\fft#1#2{{#1 \over #2}}
\def\F#1#2{{ F_{#1}^{(#2)} }}
\def\cF#1#2{{ {\cal F}_{#1}^{(#2)} }}

\def\={\, =\, }
\def\+{\, +\, }
\def\-{\, -\, }

\def\R{{\bf R}}
\def\sst#1{{\scriptscriptstyle #1}}
\def\oneone{\rlap 1\mkern4mu{\rm l}}
\def\e7{E_{7(+7)}}
\def\td{\tilde}
\def\wtd{\widetilde}
\def\im{{\rm i}}
\newcommand{\ho}[1]{$\, ^{#1}$}
\newcommand{\hoch}[1]{$\, ^{#1}$}
\newcommand{\bea}{\begin{eqnarray}}
\newcommand{\eea}{\end{eqnarray}}
\newcommand{\ra}{\rightarrow}
\newcommand{\lra}{\longrightarrow}
\newcommand{\Lra}{\Leftrightarrow}
\newcommand{\ap}{\alpha^\prime}
\newcommand{\bp}{\tilde \beta^\prime}
\newcommand{\cB}{{\cal B}}
\newcommand{\cO}{{\cal O}}
\newcommand{\vecx}{\vec{x}}
\newcommand{\vecy}{\vec{y}}
\newcommand{\vecp}{\vec{p}}
\newcommand{\vecq}{\vec{q}}
\newcommand{\tr}{{\rm tr} }
\newcommand{\Tr}{{\rm Tr} }

\newcommand{\cL}{{\cal L}}
\newcommand{\cA}{{\cal A}}
\newcommand{\cD}{{\cal D}}
\def\sst#1{{\scriptscriptstyle #1}}
\def\0{{\sst{(0)}}}
\def\1{{\sst{(1)}}}
\def\2{{\sst{(2)}}}
\def\3{{\sst{(3)}}}
\def\4{{\sst{(4)}}}
\def\5{{\sst{(5)}}}
\def\6{{\sst{(6)}}}
\def\7{{\sst{(7)}}}
\def\8{{\sst{(8)}}}
\def\ve{\varepsilon}
\def\vf{\varphi}
\def\F{\Phi}
\def\wg{\wedge}

\def \foot {\footnote}
\def \bi{\bibitem}

\def \tr {{\rm tr}}
\def \ha {{1 \over 2}}
\def \td {\tilde}
\def \ci{\cite}
\def \N {{\mathcal N}}
\def \ww {\Omega}
\def \const {{\rm const}}
\def \ss {\sum_{i=1}^3 }
\def \t {\tau}
\def\S{{\mathcal S} }
\def \nn {\nu}
\def \XX {{\rm X}}

\def \lra {\leftrightarrow}
\def \vom {{\bar \omega}}
\def \E {{\mathcal  E}} \def \J {{\mathcal  J}}
\def \YY {{\rm Y}}

\def \d {\del}
\def \rJ {{J}}
\def \sms {sigma models\ }
\def \sm {sigma model\ }
\def \L {\Lambda}
\def \gl {\ell}
\def \tr {{\rm tr\ }}
\def\z{\zeta}
\def\zi{\zeta_1}
\def\zii{\zeta_2}
\def\K{\mbox{K}}
\def\eE{\mbox{E}}   \def \vt {\vartheta}
\def \vr {\varrho}
\def \wup {w}

\def\dg{\dagger}
\def\a{\alpha}

\def\e{\varepsilon}
\def\p{\phi}
\def\ap{\alpha^\prime}
\def\I{{\cal I}}

\def\R{{\bf R}}
\def\Z{{\bf Z}}
\def\C{{\bf C}}
\def\P{{\bf P}}
\def\xb{{\bar X}}
\def\Tr{{\rm  Tr}}
\def\tr{{\rm  tr}}

\def \del{\partial}
\def\g{\gamma}
\def\s{\sigma}
\def\z{\zeta}
\def\zi{\zeta_1}
\def\zii{\zeta_2}
\def\ov{\over}

\def\I{{\cal I}}
\def\J{{\mathcal J}}
\def \ok {{1\ov \k}}
\def\LL{{\mathcal L }}
\def \jL {{J}}
\def \om {\omega}
\def \cL {{\mathcal L}} \def \cH {{\mathcal H}}
\def\E{{\mathcal E}}
\def\w{\omega}
\def\b{\beta}
\def\l{\lambda}
\def\eps{\epsilon}
\def\vep{\varepsilon}
\def \De {{\mathcal D}}

 \def \cV {{\cal V}}
\def  \Jt {  {J}_{\rm tot}    }

\def \k {\kappa}
\def\foot{\footnote}
\def \four{{\textstyle {1\ov 4}}}
 \def \third { \textstyle {1\ov 3
}}
\def\det{\hbox{det}}
\def \ci {\cite}

\def \foot {\footnote}
\def \bi{\bibitem}

\def \tr {{\rm tr}}
\def \ha {{1 \over 2}}
\def \tid {\tilde}
\def \vv {{\rm v}}
\def \tl {{\tilde \l}}
\def \XX {{\rm X}}
\def \ta {{\tilde \a}}
\def \fo { {1\ov 4}}
\def \ep {\epsilon}
\def \inti {{\int^{2\pi}_0 {d \sigma \ov 2 \pi}}}

\def \d {\partial}
\def \K {{\rm S}}
\def \el {\ell}
\def \Tr {{\rm Tr}}
\def \P {\Phi}
\def \l  {\lambda}
\def \tl {{\tilde \l}}
\def \bl {{\tilde \l}}
\def \const {{\rm const}}
\def \V {v}

\def \bv {v^*}
\def \vv {{\rm v}}
\def \LL {{\mathcal L}}
\newcommand{\PV}[1]{P_{\!\!_{V_{#1}}}}

\def \bL {\ell}
\def \M {{\mathcal M}}
\def \N {{\mathcal N}}
\def \S {{\rm S}}
\def \vn {\vec n}
\def \tl {\td \l}
\def \td {\tilde}
\def \Prod {\Pi}
\def \O {{\mathcal O}}
\def \Q {{\rm  Q}}
\def \D {\Delta}
\def \N {{\mathcal N}}
\def\tN{{\tilde N}}

\def \m {\mu}
\def \vs {\vec \s}
\def \ie {i.e.}

\def \cD {{\cal D}}

\def  \le  {\l_{\rm eff}}

\def \rS {{\rm S}}
\def\as{{\a}}
\newcommand{\bra}[1]{\mbox{$\langle #1 |$}}
\newcommand{\ket}[1]{\mbox{$| #1 \rangle$}}

\newcommand{\auth}{AUTHORS}

\def\thb{\bar{\theta}}
\def\Thb{\bar{\Theta}}
\def\barp{\bar{p}}
\def\barq{\bar{q}}
\def\barc{\bar{c}}
\def\bard{\bar{d}}
\def\e{\epsilon}

\def \bi{\bibitem}
\def \la {\label}

\def \l {\lambda}
\def\foot{\footnote}
\def \tl  {{\tilde \l}}
\def \sql {{\sqrt \l}}
\def \adss {$AdS_5 \times S^5$\ }
\newcommand{\rf}[1]{(\ref{#1})}
\def \ov {\over}

\def\th{\theta}
\def\Th{\Theta}
\def\vth{\vartheta}
\def\btheta{{\bar\theta}}
\def\ttheta{{{\tilde\theta}}}
\def\bttheta{{{\bar\ttheta}}}
\def\vth{\vartheta}
\def\vt{\vartheta}

\def\ra{\rightarrow}
\def\N{{\cal N}}
\def\F{{\cal F}}
\def\uM{\underline{M}}
\def\uN{\underline{N}}
\def\uP{\underline{P}}
\def\cc{\circ}
\def\eqv{\equiv}

\def\ni{\noindent}
\def \ha{{1\ov 2}}
\def \bw {{\rm w}}

\def\r{{\rm r}}

\def \cT {{\cal T}}
\def \no {\nonumber}

\def\a{{\rm\bf a}}
\def\b{{\rm\bf b}}
\def\c{{\rm\bf c}}

\def\Y{{\rm Y}}
\def\X{{\rm X}}
\def\tY{\tilde{\rm Y}}
\def\tX{\tilde{\rm X}}
\def\dY{\dot{\rm Y}}
\def\dX{\dot{\rm X}}

\def \J {\mathcal{J}}
\def \del {\partial}

\def\dF{\dot{F}}
\def\dG{\dot{G}}
\def\df{\dot{f}}
\def \E {{\cal E}}

\def \S {{\cal S}}
\def \J {{\cal J}}

\def\ms{\mathcal{S}}
\def\mj{\mathcal{J}}
\def\soj{\fr{\ms}{\mj}}
\def \R {{\bf R}}
\def \om {\omega}
\def \tH {\widetilde H}
\def \bE {\bar E}
\def \x {{\cal X}}

 \def \bb {\bar \beta}
\def \W {{\cal E}}

\def \bi{\bibitem}
\def \la {\label}

\def \l {\lambda}
\def\foot{\footnote}
\def \tl  {{\tilde \l}}
\def \sql {{\sqrt \l}}
\def \sqtl {{\sqrt {\tilde \l}}}

\def \HH {{\rm E}}
\def \cS {{\cal S}}
\def \cL {{\cal L}}

\def \adss {$AdS_5 \times S^5$\ }

\def \D {\Delta}
\def \thet {\theta}
 \def \t {\tau}
 \def \p {\del}
 \def \rN {{\rm N}}
 \def\tw{{\tilde w}}
 \def\hJ{{J}}
 \def\hw{{w}}
 \def\hl{{\lambda}}
 \def\hth{{\theta}}
 \def\NN{{\cal N}}
 \def \bv {{ \bar w}}
\def \vn {{\vec n}}
\newcommand{\sfrac}[2]{{\textstyle\frac{#1}{#2}}}
\def \bl {{ \bar \lambda}}

\def \ov {\over}

\def \varpi {{\rm w}}
\def \OO {{\cal O}}

\def \KK {{\rm  K}}
\def \EE {{\rm  E}}

\newcommand{\tT}{\widetilde T}
\newcommand{\tF}{\widetilde F}
\newcommand{\tG}{\widetilde G}
\newcommand{\tJ}{\widetilde J}
\newcommand{\tj}{\tilde j}
\newcommand{\arcsinh}{{\rm arcsinh}}
\newcommand{\arccosh}{\hbox{arccosh}}
\newcommand{\arctanh}{\hbox{arctanh}}
\newcommand{\sech}{\hbox{sech}}

\newcommand{\GG}{{\cal G}}
\newcommand{\CC}{{\cal C}}
\newcommand{\BB}{{\cal B}}
\newcommand{\pint}{\makebox[0pt][l]{\hspace{3.4pt}$-$}\int}
\newcommand{\up}{\uparrow}
\newcommand{\down}{\downarrow}
\newcommand{\updown}{\updownarrow}

\def \DD  {{\cal D}}
\def \SS  {{ \cal S}}
\def \sn {{\rm sn}}

\def \gt {\frac{g}{\sqrt 2}}
\def \x   {{\rm x}}

\def \te {\theta}

\def \XX {{\rm X}}
\def \rt {{\rm t}}
\def\b{\beta}
\begin{document}
\overfullrule=0pt
\parskip=2pt
\parindent=12pt
\headheight=0in \headsep=0in \topmargin=0in \oddsidemargin=0in

\vspace{ -3cm} \thispagestyle{empty} \vspace{-1cm}

\begin{flushright} HUTP-06/A0024
\end{flushright}

\begin{center}
\vspace{1.01cm}
{\Large\bf
Infinite spin limit of semiclassical  string states
\vspace{.3cm}
 }

 \vspace{.5cm} {
 J.A. Minahan$^{a,}$\footnote{On leave from Department of Theoretical Physics, Uppsala University, Uppsala Sweden},
 A. Tirziu$^{b,}$\footnote{tirziu@mps.ohio-state.edu}
 and A.A.
 Tseytlin$^{c,b,}$\footnote{Also at
 Lebedev  Institute, Moscow.
 }}\\
 \vskip 0.3cm

{\em $^{a}$Jefferson Laboratory, Harvard University\\
Cambridge, MA 02138 USA\\
\vskip 0.08cm $^{b}$Department of Physics, The Ohio State University,\\
Columbus, OH 43210, USA\\
\vskip 0.08cm $^{c}$  Blackett Laboratory,
Imperial College,  London SW7 2AZ, U.K. }

\end{center}

\vskip1cm
 \begin{abstract}
 Motivated by recent works of  Hofman and Maldacena and Dorey
  we  consider   a special infinite spin  limit
 of semiclassical  spinning
 string states in $AdS_5 \times S^5$.
 We discuss  examples of
  known folded and circular 2-spin
 string solutions  and demonstrate explicitly
  that the 1-loop superstring correction
 to the classical expression for the energy vanishes in the
 limit when  one of the spins  is much larger that the other.
 We also  give a general discussion of this  limit
 at the level of  integral  equations
 describing finite gap solutions of  the string sigma model
 and argue that the corresponding
 asymptotic form of the string and
 gauge Bethe equations is the same.

\end{abstract}
\newpage

\renewcommand{\theequation}{1.\arabic{equation}}
 \setcounter{equation}{0}

\setcounter{equation}{0} \setcounter{footnote}{0}
\setcounter{section}{0}

\section{Introduction}

In this paper, following  recent work of  \ci{beis,hm,dor1}, we
explore   a
special  limit of semiclassical  string states  in \adss
 and dual gauge theory states
in which one of the charges (one spin  $J$  in $S^5$) is much larger than  all others.
The energy (dimension) $E$ diverges with $J$  while  their difference  stays  finite.
This limit  appears to bring in remarkable  simplifications,  and thus its study
may help to further clarify the structure of the
string/gauge spectrum  of states.

If we consider for definiteness the $SU(2)$  sector or string states on $R \times S^3$
parametrized  by the  two angular  momenta $J_1,J_2$, then the limit we are interested in is, say,
 $J_2 \gg J_1 $  and $E-J_2= f(J_1, \l) + O( { 1 \ov J_2})$
 ($\l$ is the  `t Hooft coupling or the square of the  string tension).  In the semiclassical
 approximation one assumes that $\l \gg 1$   while $ \J_i = { J_i \ov \sql}$ are fixed.
 Taking this limit for a few known classical spinning string solutions \ci{gkp,ft2,ftf}
 one finds that $E-J_2$ takes a simple  ``square root'' form, and the
 analytic form of the solution  simplifies.
 This turns out to be not accidental, as these states may be  considered as bound states
 of ``giant magnons'' whose  momentum is fixed in the large  spin limit
  \ci{hm,dor1}.
  Furthermore,   their   ``square root''   dispersion
 relation appears to be exact in $\l$,  being protected by a
 residual supersymmetry in this
 limit \ci{beis,hm}.

 One of our  aims below  will be to confirm this explicitly by a
 1-loop \adss  superstring theory computation. This is a non-trivial  check
 as the presence and implications
 of the $SU(2|2) \times SU(2|2)$  (centrally extended) supersymmetry
 \ci{beis,hm}    was not yet  established directly at the level of the superstring action of
 \ci{mt}.
 We shall also supplement this  by an analysis
 of the corresponding limit of the
 gauge/string    Bethe  equations   of \ci{kmmz,bds,afs}.

 On the  dual spin chain side this  large spin   limit  corresponds
 to the large  spin chain length $J= J_1 + J_2$, and the states
 for  which $E-J$ is fixed for $J \to \infty$  are in the ``intermediate''
 part of the spin chain spectrum. For example, at the  leading 1-loop order in $\l$
 the spin chain spectrum has the following structure in the $J \to \infty$
 limit \ci{jm} (the structure of the spectrum at
 finite $\l$ is expected to be qualitatively similar):
 it starts with the ferromagnetic vacuum (BPS state) with $E-J=0$, on top of which
 come magnon states with $E-J \sim { \l \ov J^2}  + O( { 1 \ov J^3}) $ dual to BMN states,
 then come  low-energy spin wave states
 with  $E-J \sim { \l \ov J} + O( { 1 \ov J^2})$  \ci{bmsz} dual to spinning strings \ci{ft1,ft2,ts1},
 then ``intermediate'' states with $E-J \sim { \l } + O( { 1 \ov J})$ and
 finally the spinons and  the top-energy antiferromagnetic state   with
 $E-J \sim \l J + O(1) $.\foot{The ``microscopic''
 magnon states  correspond to $J_2 \gg J_1 $ with $J_1$ being finite;
 the  ``thermodynamic'' limit which was used  in  \ci{bmsz,bfst}
 to isolate the semiclassical spin wave states assumed   that  both $J_1$ and $J_2$
 are  large   but  their ratio $J_1/J_2$  (or ``filling fraction'') is  fixed.
 The present thermodynamic limit for semiclassical states
   corresponds to
    $J_2 \to \infty$ with  $1\ll J_1\ll J_2$.
      }

  While  the momenta for standard  magnon states, $ p \sim { n \ov J}$
  scale to zero with $J \to \infty$, the  momenta of  special elementary
  ``giant magnon'' states, a   finite number of
which are used to construct physical Bethe states in the
 ``intermediate'' part of the spectrum,
are fixed  in the large length limit.  The same  applies to the states  in the
 near-antiferromagnetic region which are built out of  an order $J$  number of magnons. Indeed, the
 same limit was previously  considered
in \ci{manpol} and, in particular, in \ci{rss,za,at}  in
connection with the antiferromagnetic state  of the spin chain.
The string solution counterpart of  the antiferromagnetic state
was found in \cite{rtt}.

\bigskip

Below in section 2  we shall describe the large  spin limit  of  several classical
string solutions on $S^3$  in  $S^5$ (with the corresponding
states belonging to
the $SU(2)$ sector  of the spin chain). One of them will be  new -- the
second-spin generalization of the ``giant magnon'' of \ci{hm} (independently found
recently in \ci{dor2})
while two others  will be  special cases of the known solution --
the folded spinning string of \ci{ftf,bfst} and the circular string of \ci{ft2,art}.
In all of these cases  we shall find that the  expression for the classical
energy simplifies in
 the limit $J_2 \to \infty$
 and  takes the
universal form
\be \la{gne}   E-J_2 = \sqrt{ J^2_1   + \l k^2} \    ,  \ee
where $k$ is a constant depending on a  particular solution.
The same  applies also to the circular  ($S,J)$ solution of \ci{art}   from the $SL(2)$
sector as we discuss in Appendix B.

There are  indications based on  residual supersymmetry
\ci{hm,dor1} suggesting that semiclassical string solutions
obtained in the above limit
represent  BPS states
 and thus  their  energy formula
should not receive   string $\alpha' \sim { 1 \ov \sql} $
corrections. In section 3   we shall compute the 1-loop string correction to
the energies of  folded and circular string solutions in the large
$J_2$ limit using the  methods of  \ci{ft1,ft3}. On general
grounds, the classical energy \rf{gne}  of a classical
solution may receive 1-loop string corrections of the form $E_1 =
E_1 ( \J_1), \ \ \J_1 = { J_1 \ov \sql}$. We find that  the 1-loop
correction to the energy indeed vanishes in the $\J_2 \to \infty,
\ \J_1=$fixed limit due to a nontrivial cancellation between the
contributions of the
bosonic  and fermionic fluctuation modes. This suggests (like
in the near-geodesic
  or plane-wave cases, cf. \ci{mets,bmn,pope}), that
here the superstring action expanded near  the large-spin  classical
solution has a  hidden world-sheet supersymmetry (a remnant of target-space supersymmetry after
 $\k$-symmetry gauge fixing), but so far it has not
identified  explicitly.\foot{One of the solutions for which we shall
compute the 1-loop string correction  will be the $J_1=0$   case
of the $J_2 \to \infty$ limit of the folded string solution of \ci{ftf},
which is the same as the extremal limit of the  single-spin folded string solution of \ci{gkp}.
Its  classical energy
$E-J_2 = 2 { \sql \ov \pi}$      may be viewed as a $J_1 \to 0$ limit  of
$E-J_2 = \sqrt{ J^2_1 +    { 4 \l  \ov \pi^2}} $
describing bound state of 2 giant magnons with spin  \ci{dor1}. In fact, the
 corresponding quantum state from the $SU(2)$ sector
  (i.e. the one dual to the BMN-type operator Tr$(Z... Z WZ... Z W... )$
 should have $J_1=2$, not 0. At the level of the classical solution
 (obtained within the  semiclassical expansion with  $ \l \gg 1 $
 and $\J_i = { J_i \ov \sql}$ fixed)
 one cannot of course distinguish between the $J_1=0$ and $J_1=2$ (or $J_1=$any finite number)
 cases,
 but one may question what happens  at the quantum  level.
 Assuming that the relation $E-J_2 = \sqrt{ J^2_1 +    { 4 \l  \ov \pi^2}} $
 is exact   and setting there $J_1=2$ we finish with
  $E-J_2 = 2\sqrt{ 1 +    {  \l  \ov \pi^2}} = { 2 \sql \ov \pi} +
  0 - {\pi \ov \sql}  - {\pi^3 \ov 4 (\sql)^3} + ... $.
  The absence of the 1-loop order $(\sql)^0$ correction to the $J_1=0$ solution
   is thus also consistent with this exact square root formula.
}

In section 4 we shall return to the discussion of the  large spin limit
at the classical string  level and present
the general analysis of  it using the integral  equation \ci{kmmz}
for the finite gap solutions of the  string sigma model on $S^3$.
 We shall then comment on  the infinite length limit in the
 general Bethe  ansatz equations  on the gauge  \ci{bds}  and
 the string \ci{afs,bt,hl}
 sides  and argue  that they become the same
 in this limit, i.e. the ``dressing factor''  decouples.

 In Appendix A we discuss  some technical details of the  computation
 of 1-loop correction to the energy of 2-spin folded string solution
  in the $SU(2)$ sector.

  The same large spin  limit applies also to other sectors of states and
  we illustrate this
 on the example of the $SL(2)$  sector in Appendices B and C
 and pulsating solutions in section 4.4.
In Appendix C we also consider giant
magnons in the $SL(2)$ sector.
  It turns out that these magnons have
  infinite $E-J$ as well as an infinite Lorentz spin $S$.
  This is caused by the string reaching the boundary of $AdS_5$.
   We show that there is a regularization that gives a finite
   answer and give a possible interpretation for
    this on the gauge side.

\renewcommand{\theequation}{2.\arabic{equation}}
 \setcounter{equation}{0}


\section{Large spin limit of classical     string solutions on $R \times S^3$ }

In this section we shall describe several classical string solutions
in  the infinite spin limit.
We shall consider strings  moving in   $S^3$ part of $S^5$ in \adss
  \begin{equation}\label{metric}
  ds^2=-dt^2+  d \te^2   + \cos^2\te\ d\vp_1^2   + \sin^2\te\  d\vp^2_2 \ .
  \end{equation}
  In general, a  rigid rotating string configuration that
  we are interested in may be described as a solution of Nambu action in a ``static'' gauge
  \be \la{ans}
   t= \tau \ ,\ \ \ \  \te= \te (\s) \ , \ \ \   \ \ \    \vp_1= w_1 t + \td \vp_1 ( \s) \ ,
   \ \ \ \ \   \vp_2= w_2 t + \td \vp_2 ( \s) \ ,        \ee
   and thus carries the energy $E$ and  two angular momenta $J_i\sim w_i$.

\subsection{``Giant magnons''  with  spin   }

   The ``giant magnon'' solution considered   in \ci{hm}
   was an open string with ends moving on a big circle\foot{This solution
   is a also  special case of string with spikes \ci{kru} on $S^5$ \ci{krut,ryan}.}
   which    had
   $J_1=0,\  \ E,J_2 \to \infty$  with $E-J_2= { \sql \ov \pi} \cos \theta_0$=finite.
 Here we shall generalize  it to the case of  finite non-zero $J_1$, reproducing the energy
 formula first obtained on the spin chain side as the energy
  relation for a bound state of $J_1$
 giant magnons   in \ci{dor1}\foot{We interchange notation for $J_1$ and $J_2$
 compared to \ci{dor1}.}
  \be \la{doo}
  E- J_2 = \sqrt{ J^2_1   + { \l \ov \pi^2} \sin^2 {p\ov 2 } } \ , \ \ \ \ \ \ \ \ \
  \sin {p\ov 2 } \equiv \cos \te_0    \ . \ee
  The same classical solution was independently found in \ci{dor2}
  using a  relation to the sine-Gordon model.\foot{In conformal gauge, it can
  also be obtained  as a solution of the generalized
  integrable Neumann model \ci{krt}.}
  Setting
  \be\la{dr}
   w_1= w \ ,\ \ \   w_2=1 \ ,\ \ \ \ \   \td \vp_1= - w \psi (\s)  \ ,\ \ \ \ \
  \td \vp_2=    \vp (\s) \ ,   \ee
 the Lagrangian $\LL$ of the Nambu-Goto action $\SS=\int d\tau \LL$ is then
 determined to be
 \be \la{acd} \LL= \frac{\sqrt{\lambda}}{2\pi}\int d\s\, \sqrt{ \DD  }   \ , \ee
 where
 \begin{eqnarray}\label{action}
 &&\DD=  (\dot t^2-\cos^2\theta\dot\vp_1 ^2 -  \sin^2\theta\dot \vp_2^2)\big[(\p_\s\theta)^2
 +   \cos^2\theta\,(\p_\s\vp_1)^2
+ \sin^2\theta(\p_\s\vp_2)^2\big]\nonumber\\
 &&\qquad\qquad\qquad+\ (\cos^2\theta\ \dot\vp_1 \p_\s\vp_1 + \sin^2\theta \dot\vp_2  \p_\s\vp_2)^2\nonumber
 \end{eqnarray}
or, explicitly,
\begin{equation}
\DD=\sin^2\theta(\p_\s\varphi)^2+  w^2\cos^2\theta(\p_\s\psi)^2+(1- w^2)
\cos^2\theta(\p_\s\theta)^2- w^2  \sin^2\theta\cos^2\theta\ (\p_\s\vp+\p_\s\psi)^2\,.
\end{equation}
 Varying $\LL$ with respect to $\psi$, we find the equation
\begin{equation}\label{psieq}
\frac{\p}{\p\s}\left(\frac{-\cos^4\theta\p_\s\psi+\sin^2\theta\cos^2\theta\p_\s\varphi}{\sqrt \DD }\right)=0\,,
\end{equation}
which clearly has
 \begin{equation}\label{psisol}
 \p_\s\psi=\tan^2\theta\,\p_\s\varphi
 \end{equation}
 as a special  solution.
 Substituting (\ref{psisol}) back into the action, we find
 the reduced action that determines the expression for $\theta$
 as a function of $\vp$
 \begin{eqnarray}\label{redact}
\LL&=&\frac{\sqrt{1-w^2}\sqrt{\lambda}}{2\pi}\int d\varphi\ \sqrt{r^2+ r'^2}\ , \ \ \ \ \ \ \ \ \ \
r\equiv \sin \te, \ \ \ \ \ \ r' \equiv { d r \ov d \vp}   \ .
\end{eqnarray}
 Except for the extra  $\sqrt{1-w^2}$
  prefactor, eq.(\ref{redact}) is the same expression
  found in \cite{hm}; we thus get a ``minimal'' generalization of
  the ``giant magnon'' to the case of $w \sim J_1$ non-zero.
   The explicit form of the solution for $\te$  is thus the same as in \ci{hm}
 \begin{equation}\label{xsol}
 r= \sin \theta  =\frac{\sin\theta_0}{\cos\varphi}\,, \ \ \ \
 \qquad\qquad-\frac{\pi}{2}+\theta_0\leq  \varphi\leq  \frac{\pi}{2}-\theta_0\,.
 \end{equation}
 $\te$ then varies between  $\frac{\pi}{2}$ and $\theta_0$.
 Then $L$ reduces to
 \begin{equation}\label{Lsol}
 \LL=\frac{\sqrt{\lambda}}{\pi}\  \sqrt{1-w^2}\  \sin\frac{p}{2}\,, \ \ \ \ \ \ \ \ \ \ \ \ \ \ \
 \sin \frac{p}{2}=\cos\theta_0 \ ,
 \end{equation}
 where we have assumed that the momentum $p$
 of the magnon  is related to $\theta_0$ as in \cite{hm}.

 We can then derive from (\ref{action})  and (\ref{redact}) the conserved quantities, the energy and
 the two spins,
 \begin{eqnarray}\label{cons}
 E&=&\frac{\sqrt{\lambda}}{2\pi}\int d\varphi\
 \frac{r^2+\frac{w^2r^4+r'^2}{1-w^2}}{\sqrt{(1-w^2)(r^2+r'^2)}}\nonumber\\
 J_2&=&\frac{\sqrt{\lambda}}{2\pi}\int  d\varphi \
\frac{r^2\left(\frac{w^2r^4+r'^2}{1-r^2}+w^2r^2\right)}{\sqrt{(1-w^2)(r^2+r'^2)}} \nonumber\\
J_1
&=&\frac{\sqrt{\lambda}}{2\pi} \ w \int  d\varphi\  \frac{\sqrt{r^2+r'^2}}{\sqrt{1-w^2}}
=\frac{w}{1-w^2}\,\LL\,.
\end{eqnarray}
Here $E$ and $J$ are infinite, but their difference is finite  and has the  simple form
\begin{equation}\label{EmJ}
E-J_2 =\frac{\sqrt{\lambda}}{2\pi}\int  d\varphi\ \frac{\sqrt{r^2+r'^2}}{\sqrt{1-w^2}}
=\frac{1}{1-w^2}\,\LL\,.
\end{equation}
Comparing (\ref{cons})  with (\ref{Lsol}), we find that
\begin{equation}\label{tJeq}
J_1 =\frac{w}{\sqrt{1-w^2}}\,\frac{\sqrt{\lambda}}{\pi}\sin\frac{p}{2}\,,
\end{equation}
and hence from (\ref{EmJ}) we reproduce the energy formula \rf{doo}.

To complete the solution, let us find
the dependence of  $\varphi$ on  $\psi$;
integrating (\ref{psisol}) and using (\ref{xsol}) and   (\ref{tJeq})   gives
\begin{equation}\label{psisol2}
\varphi\=\arctan\left(\cot\theta_0\tanh\left(\cot\theta_0 \ \psi\right)\right)\,.
\end{equation}
It is  also convenient   to express $\te$ in terms of $\psi$
\begin{equation}
 \te=\arccos\left(\cos\theta_0\,\sech(\cot\theta_0\ \psi)\right)  \,.
\end{equation}
At the ends of the string, $\tan\varphi=\pm\cot\theta_0$, therefore
 $\psi\to\pm\infty$.  In other words,
the string wraps infinitely many times around the $\psi$ or
 $\vp_1$  direction.
  Note that as $\theta_0\to0$, $\varphi(\psi)$ approaches the
   step function $\varphi(\psi)=\frac{\pi}{2}\epsilon(\psi)$,
    while similarly $\theta(\psi)$ approaches
     $\theta(\psi)=\frac{\pi}{2}\epsilon(\psi)$.  (We have continued $\theta$ to $\theta<0$ since $\varphi$ jumps by $\pi$ as $\psi$ changes sign).   This behavior will be relevant when considering the folded string.

  \bigskip

The  discussion of finite gap solutions in section 4 below suggests that there should exist
also more  general solutions representing  bound states of
 $n$ magnons with total momentum $p$
with energy
\be \la{ooi}
  E- J_2 = \sqrt{ J^2_1   + { \l \ov \pi^2}  n^2 \sin^2 {p\ov 2 n } }
  = n  \sqrt{  ({J_1 \ov n})^2    + { \l \ov \pi^2}  \sin^2 {p\ov 2 n } }   \ . \ee
In the case of $J_1=0$
the special case of $\te_0 =0$  or $p= \pi$  and $n=1$
corresponds to
a string that stretches through the north pole of a 2-sphere \ci{hm}. A combination of  $n=2$
of such strings  with total  $ p= 2\pi$£ and thus with
$E-J_2 =  2 { \sql \ov \pi} $  is then a limit of a folded
closed string
rotating on £$S^2$ with its center at rest at the north  pole and the positions of the folds
approaching the equator ($\theta={\pi\ov 2}$).
Similarly, there exists   an   analogous  $J_2 \to \infty$, \ $p= n \pi$   limit of
the  folded ($n\ov 2$ times)  2-spin solution
of \ci{ftf,bfst}  with the simple
energy  formula found  (for $n=2$) in \ci{dor1}\foot{The $J_2 \gg J_1 $ limit of the folded string
solution  of \ci{ftf} was discussed (for $n=2$)  in Appendix E in \ci{bfst}
where the leading term in the expansion  of the square root  at $J_1  >  \sql$ was found.}
\be \la{oo}
  E- J_2 = \sqrt{ J^2_1   + { \l \ov \pi^2}  n^2  }   \ .
  \ee
 We shall review this limit
   and present the explicit form of the resulting solution in the next
 subsection.

\subsection{ $J_2 \gg J_1$  limit of  the  folded string  solution  }

Another example is found as a limit of the
 2-spin folded string   described  in conformal
 gauge by the following ansatz
(cf. \rf{ans}, see also \ci{ts1} for a review)
\be \la{ns}
   t= \k \tau \ ,\ \ \ \  \te= \te (\s) \ , \ \ \   \ \ \    \vp_1= w_1 \tau  \ ,
   \ \ \ \ \   \vp_2= w_2 \tau  \ ,         \ee
where \ci{ftf} ($\te' \equiv \del_\sigma \te$)
\begin{equation}
\theta''+\frac{1}{2}w_{21}^2 \sin 2\theta=0, \quad \quad
w_{21}^2\equiv w_{2}^2-w_1^2 \label{thetae1}
\end{equation}
where we assumed that $w_{2}>w_1$ and for generality
introduced the scaling
parameter $\k$.\foot{When $w_2=w_1$ the
solution is $\theta=m \sigma$, where  $m$ is an integer. This is can
be transformed \ci{bfst}  into the circular rotating solution with equal spins
$J_1=J_2$. In the limit  when  $\J_{1,2}=\infty$
it has  $E=J_1+J_2$, i.e. is equivalent to a  BPS state  represented by a
point-like string.}
Then
\begin{equation}
\la{ss} \theta'^2=w_{21}^2(\sin^2 \theta_*-\sin^2 \theta) \ ,
\end{equation}
where $\theta_*$ determines the length of the folded string, i.e.
$-\theta_*\leq \theta  (\s) \leq \theta_*$.
The  conformal gauge constraint implies
\begin{equation}\la{r}
\kappa^2=\theta'^2+w_1^2 \cos^2 \theta+w_2^2 \sin^2 \theta=w_1^2
\cos^2 \theta_*+w_2^2 \sin^2 \theta_* \ .
\end{equation}
We shall consider the case of a single fold (the number of folds $n \ov 2 $
is easy to restore at any stage).
The solution of (\ref{ss}) can be written in terms of the elliptic functions
\cite{ftf,bfst}
\begin{equation}\la{rr}
\cos \theta(\sigma)={\rm dn} (  w_{21} \sigma,q), \quad \quad \sin
\theta(\sigma)=\sqrt{q} \ {\rm sn}  (w_{21}\sigma,q)  \ .
\end{equation}
\begin{equation}\la{rg}
q\equiv \sin^2 \theta_*=\frac{\kappa^2-w_1^2}{w_2^2-w_1^2} \ .
\end{equation}
The periodicity in $\sigma$ implies\foot{Here
$\KK(q)\equiv \int_{0}^{\frac{\pi}{2}}\frac{d \alpha}{\sqrt{1-q \sin^2
\alpha}}$.}   \ \
\begin{equation}\la{rgg}
2 \pi=\int_{0}^{2 \pi}d \sigma=4\int_{0}^{\theta_*}\frac{d
\theta}{w_{21} \sqrt{\sin^2 \theta_*-\sin^2 \theta}} \ , \ \ \ \ \ \ \ \ \ \
w_{21}=\frac{2}{\pi}\KK(q) \ .
\end{equation}
 The conserved charges are
\begin{equation}\la{kgg}
E=\sqrt{\lambda}\ \kappa , \quad J_1=
\sqrt{\lambda}\ w_1\int_{0}^{2\pi}\frac{d \sigma}{2 \pi}\cos^2
\theta=\frac{2\sqrt{\lambda} w_1 }{\pi
w_{21}}\int_{0}^{\theta_*}\frac{\cos^2 \theta
d\theta}{\sqrt{\sin^2 \theta_*-\sin^2 \theta}}
\end{equation}
\begin{equation}\la{lgg}
 J_2= \sqrt{\lambda}\ w_2 \int_{0}^{2\pi}\frac{d \sigma}{2 \pi}\sin^2 \theta=\frac{2\sqrt{\lambda}
w_2 }{\pi w_{21}}\int_{0}^{\theta_*}\frac{\sin^2
\theta d\theta}{\sqrt{\sin^2 \theta_*-\sin^2 \theta}} \ .
\end{equation}
 The
parameters $w_1$, $w_2$ and $\theta_*$ can be determined in terms
of $J_1$ and $J_2$ (and $\l$).\foot{Combining the above equations one obtains the two equations
that determine
${\E}={\E}({\J}_1,{\J}_2)$, where $ E=\sqrt{\lambda}\mathcal{E}  , \
J_1=\sqrt{\lambda}\mathcal{J}_1 , \ \
J_2=\sqrt{\lambda}\mathcal{J}_2$
   \cite{bfst}:
$
\big(\frac{\mathcal{E}}{\KK(q)}\big)^2-\big(\frac{\mathcal{J}_1}{\EE(q)}\big)^2=\frac{4}{\pi^2}q
\ , \ \
\big(\frac{\mathcal{J}_2}{\KK(q)-\EE(q)}\big)^2-\big(\frac{\mathcal{J}_1}
{\EE(q)}\big)^2=\frac{4}{\pi^2}.
$
}
Let us now follow \ci{bfst,dor1} and consider a  special limit of this solution
where $J_2 \gg J_1$, i.e.
 $\J_2 \to \infty$  for fixed  $\J_1$. As usual in a  semiclassical expansion we assume that
 $\l \gg 1$ and $\J_1= { J_1\ov \sql}$ is kept
finite.
It corresponds to
 the particular case when the string is maximally stretched in $\te$,
so that its angular momentum $J_2$
     around its centre  of mass
     is maximal  and goes to infinity (while the momentum of its center of mass
     $J_1$ is arbitrary).

      Let us now take the limit  $\theta_* \to {\pi\ov 2}$, i.e.
      $q\rightarrow
     1$.  Let us  distinguish two steps.  First,  the conformal constraint
     (\ref{r}) implies that  $w_2=\kappa$.  Second,
     the periodicity condition (\ref{rgg})  leads to the conclusion that
      one must have
     $w_{21}  \rightarrow \infty$. Indeed,
     in the limit $q \to 1$ we get $\KK(q)\to \infty$, so that
  \be \la{kop}
\theta_*\to \frac{\pi}{2}  \ , \ \  \ \  q\to 1 \ , \  \ \ \  {\rm i.e.} \ \    \ \ \ \ \ \ \ \ \
\ \ w_{21}, \kappa \to \infty \ .
     \ee
      If we do not impose the
     periodicity condition, we get a more
     general kink solution (see \rf{solution} below)
      which does not,
     however, represent
     a physical closed-string state.

 Setting\footnote{In order to have $\J_1$ staying finite   in the limit
 $\kappa \rightarrow \infty$ we need to rescale $w_1$.}
    \be  \la{urgg} w_1 \equiv \kappa  w \ , \ \ \ \ \ \ \
w_{21}=\kappa\sqrt{1-w^2}\ , \ \ \ \  \ \ \   w < 1 \ ,  \ee
so that  $ \vp_1=  w t, \ \vp_2 = t$, cf. \rf{dr},
we get
 from
  \rf{ss}\foot{$w=1$ thus corresponds to the BPS limit when $\te$ is constant.}
\begin{equation}
\theta' = \pm   \kappa \sqrt{1-w^2}   \cos \theta \ . \label{diff}
\end{equation}
To illustrate  what happens as $ \theta_*\to \frac{\pi}{2}$, i.e.  as    $q$ approaches 1,
one may plot the periodic solution
$\theta(\sigma)=\arcsin [\sqrt{q} \ \sn (w_{21}\sigma,q)]$
 with $\sigma$
between $-\pi$  and $\pi$ (see Fig.1).
 In the limit, $\theta (\s) $  for $-\pi   < \s < \pi$ becomes just a
 {\it step function},
 like the one considered previously,
 jumping from
 $-{\pi\ov 2}$ to $+{\pi\ov 2}$. It can then  be periodically extended  to all $\s$,
  so that $\te'\to \pm\infty $
  at $\s=-\pi, 0, \pi, ...$   and $\te'\to 0 $  at other points
  in agreement with \rf{diff}.
\begin{figure}[t]
\centerline{\includegraphics[scale=.8]{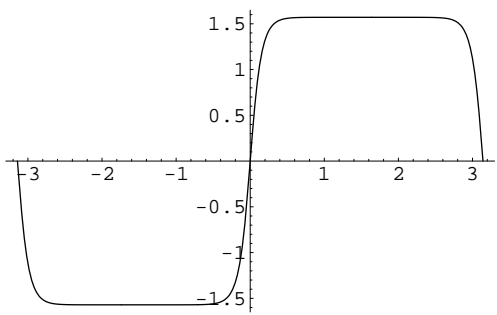}}\caption{$\theta(\sigma)=\arcsin[\sqrt{q}\ \sn(w_{21}
\sigma,q)]$, for $q=0.99999999$,  $ - \pi  \leq  \s \leq  \pi$.} \nonumber
\end{figure}
The energy  of the solution and $J_2$  then  approach infinity
\begin{equation}\la{ee}
E= \sqrt{\lambda}\ \kappa =\frac{2
\sqrt{\lambda}}{\pi \sqrt{1-w^2}}\int_{0}^{\frac{\pi}{2}}
\frac{d \te }{\cos^2 \theta} \ \to \ \infty   \ ,  \ee
\be
J_1=\frac{2 \sqrt{\lambda}}{\pi}\frac{w}{\sqrt{1-w^2}} \ ,
 \ \ \ \ \ \
\ \ \ \
J_2= \frac{2 \sqrt{\lambda}}{\pi
\sqrt{1-w^2}}\int_{0}^{\frac{\pi}{2}}d\theta\ \frac{\sin^2
\theta}{\cos^2 \theta} \ \to \ \infty \ ,
\end{equation}
 while  $E-J_2$ stays finite  \ci{dor1}
\begin{equation}\la{chu}
E-J_2= \sqrt{J_1^2+\frac{4\lambda}{\pi^2}} \ .
\end{equation}
Let us mention that  if one formally relaxes  the periodicity condition in
$\s$   and  introduces the  new  spatial variable
$x = \k \sigma \in (-\infty, \infty)$  which will be fixed in
 the limit $\k\to \infty$
then   the  solution of \rf{diff}  of the theory defined
 on a plane instead of a cylinder is
\begin{equation}
\theta(x)=\pm 2\arctan \tanh ( \ha \sqrt{ 1 - w^2} x ) \ , \ \ \ \ \ \ \  x \equiv  \k \s \ .
 \label{solution}
\end{equation}
This   non-trivial   solution \rf{solution}
(which  is  not a
limit of the  periodic solution on a circle)
appears only in the exact scaling limit
and describes  a kink localized near
 $x=0$.\foot{This solution of the sin-Gordon equation may be
 interpreted as  describing  a zero-energy
  particle that goes from one maximum of the
$- \cos^2 \theta$  potential to another  in an  infinite amount
of ``time'' $x$ (we have  $\theta(0)=0$, $\theta(x=\pm \infty)=\pm
\frac{\pi}{2}$).}

Let us mention that  for  $w=0$  eq.\rf{solution}   represents a  limit of the solution
 in \cite{hm} in the conformal gauge.
  The parameter $\te_0$  in \ci{hm}
  and in  the previous subsection
  is formally
  related to $\te_*$ by a  $\pi \ov 2$ shift. Indeed,
    here the center of the string is at
  the  pole ($\te=0$)
  and  its ends (at $\pm \te_*$) approach the equator in the limit, while in the
  previous subsection the ends of the string where at the equator
  from the start while  its center was approaching the  pole as $\theta_0\to  0$.

Another remark is that the energy formula (\ref{ooi}) suggests the existence
of  more  general closed string configurations with $J_1=0$
for which  $p=2M \pi$  with  integer $M$
 \be
  E- J_2 =   {\sqrt{\l}\  n \ov \pi} | \sin {M \pi \ov  n }  |  \ . \ee
  The  corresponding closed
string solution describes a string with spikes \ci{kru}
on $S^5$ and   was  obtained in \cite{ryan}. It has
\begin{equation}
\varphi_{1}=0, \quad \quad \varphi_{2}=\omega \tau+M \sigma, \quad
\quad \ \ \theta(\sigma)=\theta(\sigma+ 2\pi)\ .
\end{equation}
In the limit $J_2\rightarrow \infty$, one finds  that
$\omega\rightarrow 1$. For an  arbitrary winding number $M$ and
number of cusps  $n$, the closed string is built out of $n$
segments  with ends on  the $\varphi_{2}$-equator of $S^2$  (with
minimal  value of $\theta= \theta_{*}$ reached  in the middle of
each segment); all segments combine  to cover the $2 \pi M$
distance along the equator.
 For  $M=1, \ n=2$, one  recovers the folded string, or
more generally,  for $M={n\ov 2}$ one gets
 $n\ov 2$-folded string solution for which
the string stretches between the opposite  points
on the  equator  passing through the north pole in
$\theta$ (i.e. in this case $\theta_*=\frac{\pi}{2}$).

  \subsection{ $J_2 \gg J_1$  limit of  circular string solution }

 The simplest circular 2-spin string solution  on $S^3$ is represented  in
 conformal gauge  by  \ci{art} (cf. \rf{metric},\rf{ans},\rf{ns})
 \be \la{ne}
 t= \k \tau \ ,\ \ \ \  \te= \te_0={\rm const} \ , \ \ \   \ \ \    \vp_1= w_1 \tau  + m_1 \s \ ,
   \ \ \ \ \   \vp_2= w_2 \tau  + m_2 \s \ .         \ee
Written in terms  of 2 complex combinations  of embedding coordinates
of $S^3$ into $R^4$   we have
\begin{equation}
\X_{1}=a_1 e^{i w_{1}\tau+i m_1 \sigma}, \quad \X_{2}=a_2 e^{i
w_{2}\tau+i m_2 \sigma}, \quad |a_1|^2+|a_2|^2=1\ ,
\label{circ}
\end{equation}
where $a_1 = \cos \te_0, \ a_2 = \sin \te$.
The energy and two spins are
\begin{equation}
E= \sql \E, \ \ \ \ \  J_i = \sql \J_i\ ,\ \ \ \ \ \ \ \
\mathcal{E}=\kappa, \quad \mathcal{J}_i=a_i^2 w_i \ ,
\end{equation}
where the equations of motion and conformal gauge  conditions imply ($i=1,2$)
\begin{equation}\la{ko}
w_i=\sqrt{m_i^2+\nu^2}, \quad \quad
\kappa^2=2 \sum_i a^2_i w_i^2-\nu^2, \quad \quad \sum_i a_i^2 w_i m_i=0\ .
\end{equation}
This gives
\begin{equation}
\mathcal{E}^2=2 \sum_i \sqrt{m_i^2+\nu^2}\mathcal{J}_i-\nu^2, \quad \sum_im_i
\mathcal{J}_i=0, \quad\quad
\sum_i\frac{\mathcal{J}_i}{\sqrt{m_i^2+\nu^2}}=1 \ .
\end{equation}
Here we are interested in the solution
  when  $J_2\gg J_1$.  To consider this it  is useful to  fix one of the two
  winding numbers to be 1 (it is easy to restore its general
  value at the end); setting
\begin{equation}
m_2=1, \  \quad \quad m_1=-m, \quad \quad \mathcal{J}_2=m
\mathcal{J}_1
\end{equation}
we should thus expand the above relations
 in large $m$ at fixed $\J_1$.
In general, the  relation
between the spins and the energy  is found by eliminating $\nu$ from the following two equations
\begin{equation}
\frac{m
\mathcal{J}_1}{\sqrt{1+\nu^2}}+\frac{\mathcal{J}_1}{\sqrt{m^2+\nu^2}}=1,
\quad \mathcal{E}^2=2 \sqrt{1+\nu^2}m \mathcal{J}_1+2
\sqrt{m^2+\nu^2}\mathcal{J}_1-\nu^2 \ .  \label{equ}
\end{equation}
Expanding in large $m$ we get
from the first equation
\begin{equation}\la{nuu}
\nu^2=m^2 \mathcal{J}_1^2+\frac{2
\mathcal{J}_1^3}{\sqrt{1+\mathcal{J}_1^2}}m+\mathcal{J}_1^2-\frac{1+3
\mathcal{J}_1^2}{(1+\mathcal{J}_1^2)^2}+\frac{\mathcal{J}_1^3(1+6\mathcal{J}_1^2)}
{(\sqrt{1+\mathcal{J}_1^2})^7}\frac{1}{m}+O({1\ov m^2})
\end{equation}
Then the second equation in \rf{equ}   gives
\begin{equation}
\mathcal{E}=\kappa= m\mathcal{J}_1+\sqrt{1+\mathcal{J}_1^2}-
\frac{1}{2m}\frac{\mathcal{J}_1}{1+\mathcal{J}_1^2}+ O({1\ov m^2}) \ ,
\label{energy}
\end{equation}
so  that in the strict  $m \to \infty, \ \k \to \infty$ limit
we get (recalling that $\J_2= m \J_1$)
\begin{equation}
E-J_2=\sqrt{J_1^2+\lambda}     \ .  \label{er}
\end{equation}
This is similar to the   expressions \rf{doo},\rf{chu}
found  above for other solutions  in the same limit.

Let us comment on the form of the limiting solution.
In the limit the string  becomes  infinitely long (has infinite winding number $m_1$)
but  has  infinitesimal radius   and its position approaches   $\te_0= {\pi \ov 2}$.
One can formally express the limiting solution
in terms of the coordinates on $R\times R $ instead of $R \times S^1$
   which  one may keep  finite in the
limit $\k \to\infty, \  \J_2\to  \infty$. For   $m_2=1$   we get:
\begin{equation}
\X_1=a_1 e^{i  \sqrt{1+\mathcal{J}_1^{-2}}\  t\  -  \ i\mathcal{J}_1^{-1} x},
\quad \quad \X_2=a_2 e^{i t } \ , \ \ \ \ \ \ \  \ \    t= \k \tau, \ \ \   x = \k \s  \ ,
\label{circsol}
\end{equation}
where the limiting values of the parameters  $a_i$ are\foot{In general, for $m_2=1$ the  constants $a_1$, $a_2$ can be expressed as \cite{art}\
$
a_1^2=\frac{\sqrt{1+\nu^2}}{m\sqrt{m^2+\nu^2}+\sqrt{1+\nu^2}}, \ $\  $ a_2^2=\frac{m
\sqrt{m^2+\nu^2}}{m\sqrt{m^2+\nu^2}+\sqrt{1+\nu^2}}
$.}
\begin{equation}
a_1\approx \frac{\mathcal{J}_1}{ (1+\mathcal{J}_1^2)^{1/4} }\frac{1}{\sqrt{\J_2}} \to 0  ,\
\quad \quad
a_2 \approx 1-\frac{\mathcal{J}^2_1}{2\sqrt{1+\mathcal{J}_1^2}}\frac{1}{\J_2} \to 1  \ .
\end{equation}
Restoring the dependence
on the second  winding  number $m_2 \equiv k$  we get
\begin{equation}
E-J_2=\sqrt{J_1^2+\lambda  k^2 }     \ .  \label{ey}
\end{equation}
A similar  limit exists for a circular $(S,J)$
 string in the $SL(2)$
sector \ci{art};
we discuss this in Appendix B.  

\renewcommand{\theequation}{3.\arabic{equation}}
 \setcounter{equation}{0}

\section{1-loop  correction to the energy of folded and circular string
in the  $J_2 \to \infty$ limit}

In this section we shall perform a check of the
 exactness of the energy formulae for the
 folded \rf{chu} and circular \rf{er} solutions
 by computing their 1-loop string corrections
  and  showing that they vanish.   

\subsection{Folded string case}

In Appendix A we have presented some details  of the computation
of the bosonic and fermionic quadratic fluctuation actions near the
 folded string solution \rf{ns}  for arbitrary $J_1,J_2$, i.e. arbitrary  parameter
 $\te_*$.  Here we shall specialize to the  limiting  case of interest \rf{kop}:
  $\te_*=  { \pi \ov 2}, \ \k \to \infty$.

  Before  getting into the
  more technical 
  details of the  computation let us sketch
  some of its general  features. For finite $\k$  the 1-loop correction to  the energy is
  given   by  the sum over characteristic frequencies, i.e.,  symbolically,
 \be \la{hi}  E_1= { 1 \ov 2\k}  \sum^\infty_{n=-\infty}  \sum_r
   c_r \sqrt{  n^2 + M_r^2} \ , \ee
  where $c_r$ are multiplicity and  sign factors, $n$ is the discrete momentum on a circle
  $\s \in (-\pi, \pi)$ and $M_r$ are effective masses depending on parameters of the solution.
  The $1 \ov \k$   factor is the proportionality  coefficient  between the space-time and 1-d energy
  reflecting that  $t= \k \tau$. In the large $\k$ limit
  $M_r$ will scale  as $M_r\to \k \bar M_r $;  introducing
   $p_n = { n \ov \k}$ and keeping  only the leading
  order in $\k \to \infty$ one can then replace the sum over $n$ by an integral over a
  continuous momentum variable   conjugate to  spatial variable $x= \k \s$ (see also \ci{ft1}
  for a  discussion of a similar limit):
   \be \la{hii}
  E_1= \ha   \int ^\infty_{-\infty} dp \  \sum_r   c_r \sqrt{  p^2 + \bar M_r^2}  + O( {  1 \ov \k})
    \  . \ee
  The same result can be arrived  at  directly  by
  introducing  the $\k$-rescaled variables as in \rf{urgg}
 $ w_1= \k w, \   t= \k \tau, \ x = \k \s$. Then  the resulting  quadratic fluctuation action
 can  be  written as   $S= \int dt \int^\infty_{-\infty}  dx   \bar L $.

 In computing $\bar L $  and thus  $\bar M_r$  for  the present case
 of the folded solution
 we should remember to use the form of the  solution   as it
 appears in the large $\k$  limit of the original periodic   solution
 on a $\s$-circle,
 and not the formal solution on an infinite line \rf{solution}
 that exists in the strict scaling limit.
 In other words, $\te(\s)$  should be replaced by a periodic version
 of the step  function ${\pi \ov 2} \epsilon(\s)$
 which is a large $\k$ limit of the
 solution \rf{rr}.

 Let us now consider in turn the relevant bosonic  and fermionic  fluctuations
  as they appear in $\bar L$.
 The   $AdS_5$ fluctuations in \rf{fluctads}
 have rescaled  mass equal to 1,  and the masses  of two
 decoupled $S^5$ fluctuations in \rf{x3},\rf{lya} are    be given by
 \be
 \bar M^2_3= - \bar    \L = 1 - 2  (1-w^2) \cos^2 \theta  \ , \ \ \ \ \    \L= \k^2  \bar \L    \ .
 \label{x33}
\end{equation}
 The  Lagrangian for the remaining three  $S^5$ bosonic fluctuations \rf{fluct2}   takes the form
 (here $f' = \del_x f, \ \dot f = \del_t f$)
\begin{eqnarray}
\bar L=\frac{1}{2}\bigg[\dot{\eta}^2 + \dot{f}_1^2+\dot{f}_2^2   &-&\eta'^2      -f_1'^2-f_2'^2
- \bar M^2_1 ( \eta^2 + f_1^2) - \bar M^2_{2}f_2^2
\nonumber\\
&+& 4 (w \sin \theta \ f_1  -  \cos \theta\
f_2 )  \dot{\eta}
 \bigg] \ , \label{fluct3}
\end{eqnarray}
\begin{equation}
\bar M^2_{1} = (w^2-1)\cos 2\theta \ , \ \ \ \ \ \ \ \  \ \ \ \bar M^2_{2} =
(w^2-1)(1+\cos 2\theta)\ .  \label{masses}
\end{equation}
As already mentioned above,  $\theta (\s)  $  should be replaced  by
the periodic extension    of the step function
 $ {\pi \ov 2} \epsilon(\s)$   at $ - \pi  < \s < \pi$.
To leading  order in large $\k$
one  may formally replace it by  $ {\pi \ov 2} \epsilon(x)$ \foot{
More precisely,  one needs also  to include step functions at $\pm \infty$.
 It turns out that contributions  of
   isolated points, such as
 $x=0,\pm \infty$,  may  be ignored
  when computing the spectrum.}
\be \theta(x) = {\pi \ov 2} \epsilon(x) \ , \ \ \ \ \ \
\ \ \  \ \ \epsilon(x)=\bigg\{\begin{array}{ll}
    1, & x  > 0 \\
    0, & x=0\\
    - 1, & x  <  0 \\
\end{array}
\la{eps}\ee
Thus  $\theta$ is essentially constant for $x >0$ and for
$x <0$  (i.e. the string  is close to a point-like geodesic
state).\foot{This  limit  of the folded solution written in  cartesian
coordinates is
$\X_1=[1-\epsilon^2(x) ]e^{iw t}, \ \ \X_2=\epsilon(x)  e^{i t}$,
so that the size of the string shrinks to zero
in $\X_1$ plane  apart from $x=0$ (and $x=\pm \infty$).
This is similar to what was found  in the case of the circular
solution (\ref{circsol}).}
Then   \be \sin \te = \epsilon(x) \ ,\
\ \ \ \ \ \ \ \ \cos \te  = 1 - \epsilon^2(x) = \bigg\{
\begin{array}{ll}
    0, & x \not=0 \\
    1, & x=0\\
\end{array}%
 \ .
\ee
If  we  ignore the contribution of the
point $x =0$, \footnote{A qualitative reason  why  one can ignore
the contribution of this
 single point is that we are computing an extensive quantity
 and the coefficient function  in the corresponding
 differential equation for the fluctuations is finite at
  this point (i.e.
 this is different from, e.g., a delta-function  potential case).}
  we find that the mass \rf{x33} of the two decoupled $S^5$  
fluctuations  becomes equal  to 1, and that $f_2$ in \rf{fluct3}
becomes massless and decouples.
We are  left with
 the following  Lagrangian  for $\eta$ and $f_1$
\begin{eqnarray}
\bar L=\frac{1}{2}\bigg[     \dot{\eta}^2-\eta'^2  +      \dot{f}_1^2-f_1'^2
+ (w^2-1)  (\eta^2 +  f_1^2) \ + \  4 w \epsilon(x)  f_1 \dot{\eta}
 \bigg]  \ . \label{fluct4}
\end{eqnarray}
Using that   $\epsilon^2 = 1$ away from the point $x=0$, 
  we end up with the following
characteristic frequencies (conjugate to time variable $t$)
\begin{equation}
\omega=\pm  w \pm \sqrt{p^2+1} \label{cfreq} \ ,
\end{equation}
where   $p$ is a  continuous 1-dimensional momentum corresponding
to  the $x$-direction. We explain   the derivation of \rf{cfreq}
in detail at the end of Appendix A.

\bigskip

Let us now  consider the fermionic fluctuations in \rf{fe},\rf{Ffluct1},
where we set $w_2= \k$ and rescale  the coordinates  by $\k$. 
We shall also use that \rf{diff} implies
$\theta'=\pm
\sqrt{1-w^2}\cos \theta$,\ $\theta''=-(1-w^2) \sin \theta
\cos \theta$, where  here and below prime stands for $\del_x$ (and
dot for $\del_t$).
 To simplify the fermionic operator $D_F$ in
(\ref{Ffluct1}) we perform  the rotations in the $(89)$ and
$(08)$-planes:
\bea \vt= e^{\fr12
s\G_8\G_9}e^{\fr12 v \G_0\G_8} \ \tilde{\vt} , \quad\quad
 \sin s=\fr{1}{u} \sin \theta\ , \quad \cos s=\fr{w}{u} \cos \theta \ ,  \label{spcp}
 \eea
\begin{equation}
u= \tanh v= \sqrt{\sin^2 \theta+w^2 \cos^2 \theta},\quad  \quad \cosh v=
\frac{1}{\sqrt{1-w^2}\cos \theta} \ .
\end{equation}
Then $D_F$ becomes
\begin{eqnarray}
D_F&=& \Gamma_0 \sqrt{1-w^2}\cos \theta
\partial_{t}-\theta' \Gamma_7 \partial_{x}+\ u  \theta'
\Gamma_{078}
\Gamma_{1234}\nonumber\\
&+&\frac{ \sqrt{1-w^2}}{2 u}\Gamma_0 \cos
\theta\bigg[\sqrt{1-w^2}\sin \theta (-u \Gamma_0+\Gamma_8)+w
\Gamma_9\bigg]\Gamma_7\\
&+& \theta'^2 ( \frac{1}{2u} \tan \te \ \Gamma_{0} -\frac{w}{2u^2
\sqrt{1-w^2}\cos \theta}\Gamma_{9}) \G_{78} \ +\theta'^2
\frac{w}{2 u \sqrt{1-w^2}\cos \theta} \Gamma_{709}\ \nonumber.
\end{eqnarray}
If we further do a rescaling of the fermionic variable, introducing 
\begin{equation}
\Theta=\sqrt{\theta'}\  \tilde{\vt } \ ,\ \ \ \ \ \ \ \ \ \ \
 L_F=-2 i \kappa \bar{\Theta} \hat{D}_F \Theta\ ,
\end{equation}
we
obtain
\begin{eqnarray}
\hat{D}_F&=&\pm \Gamma_0\partial_{t}-\Gamma_7 \partial_{x}\mp
\frac{w}{2u^2} \Gamma_{789} +u \Gamma_{078}\Gamma_{1234}
\label{Ffluct2}\ ,
\end{eqnarray}
where the upper signs correspond to $x<0$, while the lower
 sign to  
$x>0$  (they come from $\del_x \theta=\pm  \sqrt{1-w^2}\cos
\theta$).
Since
$\Gamma_{1234}^2=1$, we can restrict to subspaces  satisfying
$\Gamma_{1234}\Theta=\pm \Theta$.

Let us now specialize to the relevant case when $\te$  is replaced by the
  step-function \rf{eps}. Ignoring again the contribution of the $x=0$ point
and using  that  then $u=1$ for $x<0$, and $u=-1$ for $x>0$, we get
 \begin{eqnarray}
\hat{D}_F&=&\pm \Gamma_0\partial_{t}-\Gamma_7 \partial_{x}\mp
\frac{w}{2} \Gamma_{789} \pm  \Gamma_{078} \ . \label{uct}\
\end{eqnarray}
 Computing the determinant of this  operator (now having constant
 coefficients), and solving the resulting characteristic
 equations  on either side of $x=0$,
 one finds  that the corresponding  frequencies
 are  similar to \rf{cfreq}, i.e. the are essentially the BMN ones up to a  $w$-dependent shift,
 \begin{equation}
\omega=\pm   \frac{w}{2} \pm \sqrt{p^2+1} \label{req} \ .
\end{equation}
 Combining the contributions of all  modes
 to the 1-loop shift of the energy (taking into account  proper sign factors in  \rf{e1}
 implying that the $w$-dependent shifts in \rf{cfreq} and \rf{req} drop out)
 one  finds that, just as in the BMN case,
 the   8 non-trivial bosonic  mode contributions
 cancel against  the 8 fermionic contributions, therefore, the  
 1-loop correction to the energy vanishes,
 \be  E_1=0    \ . \ee

\subsection{Circular string case  }

Let us now  perform  a similar computation in the case
of the large spin limit of the circular solution
discussed in section 2.3.
The bosonic fluctuation Lagrangian near the  circular solution
with generic $\J_1,\J_2$  was
found in \cite{art}.
In addition to 4 $AdS_5$    massive fluctuations
with mass $\k$
 there are $2$ free  fluctuations (corresponding
to the $\X_{3}$ direction of $S^5$)  which have  mass $\nu$.
Using \rf{nuu}   and rescaling the  coordinates by $\k$
 as above, we  end up with  the corresponding  characteristic frequencies,
 given in the $\k\to \infty$ limit by the same expression
\begin{equation}\la{fgh}
\omega=\pm \sqrt{p^2+  1 }  \ .
\end{equation}
The remaining 3 coupled $S^5$ fluctuations in general
are described  by  the following
Lagrangian \cite{art}
\begin{equation}
L=\frac{1}{2}(\dot{f}_1^2+ \dot{f}_2^2+\dot{g}_2^2-f_1'^2-
f'^2_2   -g'^2_2 )+ 2(a_2 w_1 f_1-a_1 w_2 f_2)\dot{g}_2-2(a_2 m_1
f_1-a_1 m_2 f_2)g_2'\label{boseq1}
\end{equation}
 Setting  $m_2=1$, $m_1=-m$  and  rescaling the world-sheet coordinates
 by $\k= m \J_1 \to \infty $  (see \rf{energy}) we  end up with the following analog of
 \rf{fluct3}
\begin{equation}
\bar L= \bigg[\frac{1}{2}(\dot{f}_1^2+
\dot{f}_2^2+\dot{g}_2^2      -f_1'^2-
f'^2_2  -g'^2_2          )+
2 \dot{g}_2 f_1 \sqrt{1+\g^2  }+2\g  g_2' f_1\bigg]      \ ,
       \label{boseq}
\end{equation}
\be \la{ga}   \g \equiv    \J_1^{-1} \ . \ee
 $f_2$ thus decouples  in the limit and becomes
 massless.
 The non-trivial  characteristic
frequencies are then found to be
\begin{equation}
\omega_{1,2}= \sqrt{1+\g^2}   \pm   \sqrt{(p+ \g )^2   + 1 }  \ , \ \ \ \ \ \ \
\omega_{3,4}= -\sqrt{1+\g^2}   \pm   \sqrt{(p- \g )^2   + 1 } \label{w13} \ .
\end{equation}
Interestingly, while the circular  solution  is unstable at finite $\J_2$
\ci{art},   it becomes stable in the present limit, i.e.
all characteristic frequencies are real.

The  fermionic fluctuation Lagrangian for the general
circular solution with two
unequal spins  was found  in
\cite{fs} (see also \ci{ryang}).
In the notation of
\cite{fs}
\begin{equation}
L=2i \ \bar \vt  D_{F}  \vt , \quad \quad \ \ \ \
  D_F=\bigg(\begin{array}{cc}
  \Delta^{+}_F & 0 \\
  0 &  \Delta^{-}_F \ , \\
\end{array}\bigg)\otimes 1
\end{equation}
\begin{equation}
\Delta^{\pm}_F=\bar{\sigma}^{a}\partial_{a}\mp W
\bar{\sigma}^{012}\mp Q \bar{\sigma}^{134}\ ,
\end{equation}
where $\bar{\sigma}^{\mu}$, $\sigma^{\mu}$ are $16\times 16$ gamma
matrices in ten dimensions and $a=0,1$.
 Here
\begin{equation}
W^2=a_1^2 (m_1^2+\nu^2)+a_2^2(m_2^2+\nu^2), \quad M^2=a_1^2
m_1^2+a_2^2 m_2^2, \quad  Q=
\frac{a_1a_2}{2MW}\kappa (m_1^2-m_2^2).
\end{equation}
One can compute the characteristic frequencies from the following
determinant
\begin{equation}
\det
\Delta^{\pm}_F=(\partial_{0}^2-\partial_{1}^2)^2
+2W^2(\partial_{0}^2-\partial_{1}^2)+
2 Q^2
(\partial_{0}^2+\partial_{1}^2)+(Q^2+W^2)^2 =0\ ,
\end{equation}
In the large $\k$ limit one finds
\begin{equation}
W^2=\k^2 +..., \quad
M^2=\frac{\k \g }{\sqrt{1+\g^2}}+..., \quad
Q^2=\frac{1}{4} \g \k +...
\end{equation}
After the rescaling of world-sheet coordinates we get from
 $\det \Delta^{\pm}_F=0$  the following fermionic
 characteristic frequencies (with 4-fold degeneracy)
\begin{equation}
\omega  =\pm \sqrt{  \big(p\pm
\frac{1}{2}\g \big)^2 + 1 } \ .
\end{equation}
Collecting  the  resulting bosonic and fermionic frequencies
and observing that after the rescaling of $\tau$  by $\k$
the 2d and space-time energies are the same, we  finish with
the following expression for the 1-loop correction to the energy\foot{
Upon using the signs factors from  (\ref{signs}) for the
contributions of the  frequencies
(\ref{w13}) one finds that the $p$-independent parts of them
cancel out.}
\begin{eqnarray}
E_1=\frac{1}{2}\int_{-\infty}^{\infty} dp
&\bigg[&  6\sqrt{p^2+1}+\sqrt{(p+\g )^2+1}+
\sqrt{(p- \g  )^2+ 1 }\nonumber\\
&-& \ 4 \sqrt{\big(p+\frac{1}{2} \g \big)^2+ 1}-4
\sqrt{\big(p-\frac{1}{2}\g\big)^2+1}\ \bigg]  \  .
\label{op}
\end{eqnarray}
This  integral is convergent,
and evaluating it directly one finds that it vanishes,
\begin{equation}
E_1=0 \ .
\end{equation}
It is interesting to note that this vanishing is due
to a non-trivial cancellation between the fermionic and bosonic contributions.
Indeed, if we shift the fermions momentum in (\ref{op}) by $r$, the resulting integral is still convergent,
\begin{eqnarray}\la{jjj}
I(\g,r)\equiv \frac{1}{2}\int_{-\infty}^{\infty} dp
&\bigg[&  6\sqrt{p^2+1}+\sqrt{(p+ \g )^2+1}+
\sqrt{(p- \g  )^2+ 1 }\nonumber\\
&-& \ 4 \sqrt{\big(p+  r \big)^2+ 1}-4 \sqrt{\big(p-  r\big)^2+1}\ \bigg]  =
\g^2-4 r^2 \ .
\end{eqnarray}
However,  it vanishes only if $r= \ha \g$  as in \rf{op},
 suggesting
the presence of  hidden $2d$ supersymmetry in this problem.

The  generalization of the above expressions
to the   case of  non-trivial  second winding number $ m_2=k$
can be  found   by replacing
  $\g= \J_1^{-1} \rightarrow  k \J_1^{-1}$;
 this does not  
 change the conclusion about
  the vanishing of the 1-loop correction
 to the  energy  in this limit.


\bigskip

\renewcommand{\theequation}{4.\arabic{equation}}
 \setcounter{equation}{0}

\section{Infinite spin limit and bound magnons
 in  integral Bethe equations}

In \cite{kmmz} it was shown how to generate classical
solutions for strings propagating on $R\times S^3$ and compare
the results to gauge theory predictions using finite gap equations.
In this section we will discuss the scaling limit and solutions of
 \ci{hm,dor1,dor2}  and section 2
 using this formalism.

 This will then allow us, in particular,  to
  argue  that  gauge theory   and string theory predictions
  should match in this limit.

\subsection{Classical finite gap  equations  for a string on $R \times S^3$ }

Let us  first summarize the results of  \cite{kmmz}.
The string sigma model
action on $R\times S^3$ in conformal gauge can be written as
\begin{equation}
\label{cf}
S= -\frac{\sqrt{\lambda}}{4\pi}
\int d\tau d\sigma\, \left[ - (\p_a t)^2  +  \frac{1}{2}\,{\rm Tr} (
j_a^2)\right] \  ,
\end{equation}
where  $j_a$   are the right currents which are written in terms of the $SU(2)$ group element $\GG$ as
$j_a=\GG^{-1}\p_a \GG={1\over 2i}j^A_a\sigma^A$.  The equations of motion that follow from (\ref{cf}) are
\be\label{sm/eq}
\p_+j_-+\p_-j_+=0, \ \ \ \ \  \ \  \ \ \
\p_+j_--\p_-j_++[j_+,j_-]=0 \ , \ \ \ \ \ \ \p_+\p_- t=0, \ .
\ee
We can also define  the left currents  $l_a=\GG j_a\GG^{-1}=\partial_a \GG\,\GG^{-1}$.
 The charges coming from the
 third component of the  left and right currents are
\be\la{qq}
Q_L^3= \frac{\sqrt{\lambda}}{4\pi}\int d\sigma\, l^3_0=J_2+J_1\ ,\  \  \ \ \ \ \ \ \ \ \
Q_R^3= \frac{\sqrt{\lambda}}{4\pi}\int d\sigma\, j^3_0=J_2-J_1\ .
\ee
A solution for $t$ in (\ref{sm/eq}) is $t=\kappa\tau$, and so the string energy $E$ is given by
\begin{equation}
E=\frac{\sqrt{\lambda}}{2\pi}\int_0^{2\pi} d\sigma\, \p_\tau  t
=\sqrt{\lambda}\,\kappa\,.
\end{equation}
We can now set up a pair of linear equations that are satisfied provided the string equations
of motion are satisfied:
\begin{eqnarray}\label{laxcf}
&&
\bigg[\p_\sigma + \gt  \bigg({j_+\over \gt -\x} - {j_-\over \gt+\x}\bigg)\bigg]\Psi
=0\,,\nonumber
\\
&&
\bigg[\p_\tau + {2\pi}\gt \bigg({j_+\over\gt -\x} + {j_-\over \gt   +\x}\bigg)\bigg]\Psi
 =0\,, \ \ \ \ \ \ \ \ \ g^2\equiv\frac{\lambda}{8\pi^2} \ .
\end{eqnarray}
where $\x$ is a spectral parameter (not to be confused with
the  spatial coordinate used
in the previous sections).
  The first equation can be integrated to give
   the monodromy matrix (given by path-ordered  product)
\be\label{monmat}
\Omega(\x)={\rm P}\exp\int_0^{2\pi} d\sigma\,
 \frac{g}{2\sqrt{2}}\left( {j_+\over \gt -\x} -
 {j_-\over  \gt +\x}\right)\ .
\ee
Because of its  unimodularity $\Omega(\x)$
 has eigenvalues $e^{\pm i P(x)}$ and satisfies the equation
\be
\Tr\Omega(\x)=2\cos P(\x)\,,
\ee
where $P(\x)$ is the quasi-momentum.
It is clear from  the Virasoro constraints
\be
\label{vircf}
\ha{\rm Tr} j_+^2=\ha{\rm Tr} j_-^2=-\kappa^2\,,
\ee
 and  (\ref{monmat})  that $P(\x)$ has the pole structure
\be\label{pres1}
P(\x)=-\frac{E/4}{\x\pm \gt }+\ldots
~~~~~(\x\rightarrow\mp \gt).
\ee
The asymptotic properties of $P(\x)$ are determined by the charges $Q_L$ and $Q_R$.  For large $\x$,
$P(\x)$ behaves as
\be\label{xtoinf}
P(\x)=-\frac{J_2-J_1}{2\x}+\ldots~~~~~(\x\rightarrow\infty)\,.
\ee
For small $\x$, using $\Omega(0)=1$ and expanding about $\x=0$, one finds
\be\label{xto0}
P(\x)=2\pi m+\frac{J_2+J_1}{2 }\, \x+\ldots ~~~~~(\x\rightarrow 0)\,.
\ee
Here  $m$ is an integer, which
follows from the periodicity condition in $\s$ for
a closed string.
We will refer to $2\pi m$ as
the string momentum, and this  can be
 thought of as a level matching condition on the string.

Since  $P(\x)$ is not single valued, there can be an interesting singularity structure in the $\x$ complex plane.
 There are two types of singularities that we can have.  First, there can be branch cuts along contours
 $\CC_k$ where  two eigenvalues of the monodromy matrix are interchanged on either side of the cut, up
  to a factor of $2\pi$.  Hence,
\be\label{cutcond}
P(\x+i0)+P(\x-i0)=2\pi n_k,~~~~~\x\in {\CC}_k\,.
\ee
We can also have singular points in the complex plane such that $P(\x)$ jumps by a multiple
 of $2\pi$ when transported around the singularity.  These singularities will pair
  up such that $P(\x)$ jumps by a multiple of $2\pi$ when it crosses a contour between the
  two singularities.  We call this contour a condensate and  label condensate $j$  by $\BB_j$.

Because of the cuts $\CC_k$, the spectral parameter space becomes
 a two-sheeted surface, with the singularities
 in (\ref{pres1}) appearing on both sheets.  It is convenient to define the resolvent $G(\x)$
\begin{equation}
G(\x)=P(\x)+\frac{E/4}{\x+ \gt }+\frac{E/4}{\x- \gt } \ ,
\end{equation}
which is free of these poles on the top sheet.
 Hence, on this  {\it physical} sheet, $G(\x)$ can be expressed as
\be
G(\x)=\sum_k\int_{\CC_k}d\x'\frac{\rho(\x')}{\x-\x'}+\sum_j\int_{\BB_j}d\x'\frac{\rho(\x')}{\x-\x'}\,,
\ee
where $\rho(\x')$ acts as a density along the cuts and condensates.
 The density along a condensate is readily determined to be $\rho(\x')=-i\,n_j$ if $\x'\in\BB_j$.
  Along the cuts, the condition in (\ref{cutcond}) can be reformulated as an integral equation for the density
\be\label{cbethe}
G(\x+i0)+G(\x-i0) =
2\pint d\x'\,\frac{\rho(\x')}{\x-\x'}=\frac{\x E}{\x^2-g^2/2}
+2\pi n_k, \qquad \x\in \CC_k.
\ee
The asymptotic behavior for large and small $\x$ in
 (\ref{xtoinf}) and (\ref{xto0})  leads to the conditions
\begin{eqnarray}\label{rhocond}
\int d\x\, \rho(\x)&=&J_1+\frac{E-J_2-J_1}{2}\ ,\nonumber\\
\int d\x\,\frac{\rho(\x)}{\x}  &=&  2\pi m\ , \nonumber\\
\int d\x\,\frac{\rho(\x)}{\x^2} &=& \frac{E-J_2-J_1}{g^2} \ .
\end{eqnarray}
We can then rewrite the integral equation in (\ref{cbethe})  in
 terms of the inputs $J_1$ and $J_2$  as
\be\label{sinteq}
2\pint d\x'\,\frac{\rho(\x')}{\x-\x'}=\frac{\x(J_1+J_2)}{\x^2-g^2/2}+g^2 \x\int d\x'\,\frac{\rho(\x')}{{\x'}^2(\x^2-g^2/2)}
+2\pi n_k, \qquad \x\in \CC_k.
\ee
This integral equation \ci{kmmz} is normally the main tool for finding string solutions, but we will see that it is
 not relevant for solutions made up only of  giant magnons!

 \bigskip

\subsection{Infinite  $J$   limit   and  matching  to  asymptotic
spin chain Bethe  equations }

Eq.\rf{sinteq}
can   be  compared to the  integral  equation
that  follows in the ``thermodynamic''  ($J_1, J_2  \gg 1 ,
 \  { J_1 \ov J_2}
=$fixed)  limit  \ci{ft2,bmsz}
from the proposed asymptotic Bethe  ansatz  on the gauge theory side \ci{bds}
\be\label{ginteq}
2\pint d\x'\,\frac{\rho(\x')}{\x-\x'}=\frac{\x(J_1+J_2)}{\x^2-g^2/2}+g^2
 \x\int d\x'\,\frac{\rho(\x')}{{\x\x'}(\x\x'-g^2/2)}
+2\pi n_k, \qquad \x\in \CC_k \ .
\ee
In general, the two equations \rf{sinteq}  and  \rf{ginteq}   do not match starting with ``3-loop'' order
implying the need
  to introduce an extra ``dressing factor''
into the spin chain Bethe ansatz \ci{afs}.

If we now  consider the scaling
 limit in which  $E$ and $J_2$ become  infinite, but
their difference $E-J_2$
 as well as  $J_1$  stay  finite, then
 it follows from (\ref{rhocond}) that the
  second term in the r.h.s. of (\ref{sinteq})
  which was the cause of
   difference between  (\ref{sinteq})  and  \rf{ginteq}
     is vanishingly small
   compared to the first term.  This also implies that
   the l.h.s of (\ref{sinteq}) is negligible,  and hence
    $n_k$  must be  infinite. As a result,
 the cut  must have  shrunk to a point.

 In general, the  above integral equations should  receive also
 contributions from   string loop corrections \ci{bt,hl}.
 The 1-loop correction  to the dressing  phase considered in \ci{hl} produces
 extra  contributions to the r.h.s. of the integral Bethe equation \rf{ginteq}, but
 it is easy to see  (e.g., from eq.(10) in  \ci{hl})
 that it
  is negligible  in the present  limit. This
 implies that the predictions  of the asymptotic
 ``undressed'' gauge theory Bethe
 ansatz  of \ci{bds}    and  full string
 Bethe ansatz   should agree in this  limit.\foot{
 This  conclusion is  consistent with  the discussion in \ci{dor1}
 where bound states  of ginat magnons   where interpreted as
 poles of BDS $S$-matrix; it was assumed that the dressing
 factor does not introduce new poles.
 }

\subsection{Giant magnons  and their bound states as finite-gap solutions }

Let  us now  consider some  simple solutions of   equations
 \rf{rhocond} and \rf{sinteq}  in the infinite $J_2$ limit.
We start with solutions made up only of condensates and  no cuts $\CC_k$.   Without cuts we can disregard
  eq.(\ref{sinteq}) and the condensates, whose contribution  to the energy, spins and string momentum is additive, can be treated individually.
The periodicity of the closed string forces the total string momentum to be an integer multiple of $2\pi$.  However, the momentum $p$ from an individual condensate need not
satisfy this condition as  long as
 the total momentum
 coming from  all the condensates that make up the closed string solution does satisfy the condition.

Hence, we may formally
 consider the case of a single condensate only, remembering
that the final
physical closed string solution
will be made up of more than
 one condensate.
  It is useful to introduce  a  different
   spectral parameter
    $y$ which satisfies $y= \x+ {g^2 \ov 2\x}$ \cite{bds}.  Then
    the equations on $\rho$ in (\ref{rhocond}) become
\begin{eqnarray}\label{rhocondy}
\int_\BB dy\, \rho(y)&=&J_1 \ ,\nonumber\\
\int_\BB dy\,\frac{\rho(y)}{\sqrt{y^2-2g^2}}  &=&p \ , \nonumber\\
2g^2\int_\BB dy\,\frac{\rho(y)}{y\sqrt{y^2-2g^2}+y^2-2g^2}&=&{E-J_2-J_1}\,.
\end{eqnarray}
In order for the momentum and the energy to be real we also require that the
 end points of the condensate be complex conjugate to each other.
  Assuming that $\rho(y)=-i\,n$, we see from the first equation in (\ref{rhocondy})
   that the end points of the condensate are $y_0\pm iJ_1/2$, where $y_0$ is to be
    determined.  If we interpret $\rho(y)$ as a density of Bethe roots, then the contour
     would naturally be chosen to be a straight line along the imaginary direction in order
     that $dy\rho(y)$ is positive real.  However, because of the square root in the second
     and third integral equations, there is a branch cut between
     $\pm \sqrt{2}g$ and so
      there is an ambiguity in how one chooses the contour.
       In particular, if we substitute
      this density  into the second equation, we find the relation
\begin{equation}\label{iemom}
\arccosh\left(\frac{y_0+iJ_1/2}{\sqrt{2}g}\right)
-\arccosh\left(\frac{y_0-iJ_1/2}{\sqrt{2}g}
\right)=i\, \frac{p}{n}\ ,   
\end{equation}
where one can see a sign ambiguity in evaluating the arccosh.
If we momentarily set $J_1=0$, then one can have
the solution $y_0=\sqrt{2} g \cos\frac{p}{2n}$, assuming
that the end points are evaluated on opposite sides of
the cut, which requires the contour to go outside one of
 the branch points.  Otherwise, there is a solution only
  if ${p}/{n}$ is a multiple of $2\pi$.  The more general solution is
\begin{equation}\label{y0eq}
y_0=\sqrt{2g^2\cos^2\frac{p}{2n}+\left(\frac{J_1}{2n}\right)^2\cot^2\frac{p}{2n}}\,
\  ,
\end{equation}
where one finds that a straight-line contour is possible if
\begin{equation}
J_1\   >\ 2\sqrt{2}\, g\, n \ \sin \frac{p}{2n}\,\tan \frac{p}{2n}\,.
\end{equation}
If we start with $J_1$ satisfying this bound and smoothly
decrease the value, one will see that the contour
starts deforming once $J_1$ is less than the bound.
 Even as $J_1\to0$, we are still left with a nontrivial contour.  This is demonstrated in figure 2 where we show three contours with different values of $J_1$ and fixed $p$.
\begin{figure}[t]
\centerline{\includegraphics[scale=1.0]{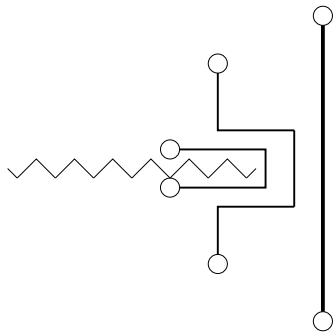}}\caption{Condensates
for three different values of $J_1$.  As $J_1\to0$, the end points
 of the contour approach the cut.} \nonumber
\end{figure}

Finally, performing the final integral and putting in
the value for $y_0$ in (\ref{y0eq}), one finds
\begin{equation}\label{E-J_2}
E-J_2=n\sqrt{\left(\frac{J_1}{n}\right)^2+8g^2\,\sin^2\frac{p}{2n}}\,,
\end{equation}
which is the same as \rf{ooi}.
The case with $n=1$  corresponds to
a single magnon with spin \ci{dor1}.
  Other values of $n$ represent bound  states  of
   $n$  magnons with the string momentum $p$ and $S^3$ angular
   momentum shared equally among magnons.

   \bigskip

We can also derive similar relations directly from the discrete BDS Bethe equations.
  These equations for the Bethe roots $y_j$  are \cite{bds}
\begin{equation}\label{BDSbethe}
\left(\frac{\x(y_j+i/2)}{\x(y_j-i/2)}\right)^{J_1+J_2}=\prod_{k\ne j}^{J_1}
\frac{y_j-y_k+i}{y_j-y_k-i}\,
\end{equation}
where $\x(y)=(y+\sqrt{y^2-2g^2})/2$.  In the limit where
  $J_2\to\infty$, there can be Bethe string solutions,
  where a string is made up of $J_1$ roots situated at
  $y_j=y_0+i(J_1+1-2j)/2$ with $j=1,...,J_1$ and $y_0$ real.  The momentum contribution of a root satisfies
\begin{equation}
e^{ip_j}=\frac{\x(y_j+i/2)}{\x(y_j-i/2)}\ ,
\end{equation}
and so the total momentum coming from a Bethe string is
\begin{eqnarray}
i\,p&=&\sum_{j=1}^{J_1}\Big[\ln[\x(y_0+i(J_1+2-2j)/2)]-\ln[\x(y_0+i
(J_1-2j)/2)]\Big]\nonumber\\
&=&\ln\big[\x(y_0+iJ_1/2)\big]-\ln\big[\x(y_0-iJ_1/2)\big]\nonumber\\
&=&\arccosh\left(\frac{y_0+iJ_1/2}{\sqrt{2}\,g}\right)-
\arccosh\left(\frac{y_0-iJ_1/2}{\sqrt{2}\,g}\right)\ ,
\end{eqnarray}
which matches \rf{iemom} when $n=1$.
Likewise, $E-J_2-J_1$ is \ci{bds}
\begin{eqnarray}
E-J_2-J_1&=&ig^2\sum_{j=1}^{J_1}\left(\frac{1}{\x(y_j+i/2)}-\frac{1}{\x(y_j-i/2)}\right)\nonumber\\
&=&ig^2\left(\frac{1}{\x(y_0+iJ_1/2)}-\frac{1}{\x(y_0-iJ_1/2)}\right)\,.
\end{eqnarray}
It is straightforward to show that this is the result for the
third integral in \rf{rhocondy} when $\rho=-i$; thus we get
 \rf{E-J_2}  with $n=1$.  More general values of $n$
 are obtained by increasing the density of the roots.

In Appendix C we will derive an analogous equation for the $SL(2)$ sector.  The other
rank one sector, the $SU(1|1)$ sector, which is equivalent
to free fermions in the one-loop approximation,
 does not have the poles and zeros in its
 S-matrix \ci{Staudacher:2004tk,Beisert:2005fw}  to
 build up (bound states of) giant magnons.


\subsection{Finite-gap solutions for large
 spin limits of circular and pulsating strings }

\bigskip

One interesting application  of this discussion
is a limit of the circular string solution of  \cite{art} considered already in section 2.3.
 Here we have  $nJ_1=mJ_2$, so that
  $J_2\to\infty$ with finite $J_1$ corresponds to holding $m$ fixed as $n\to\infty$.
   In \cite{bmsz,kmmz} it was argued that these solutions correspond to single-cut configurations
   and so $G(\x)$  is an algebraic function
\begin{eqnarray}\label{Geq}
G(\x)&=&\frac{L}{4}\bigg(\frac{1}{\x-\gt}+\frac{1}{\x+\gt}\bigg)\\
&&\ \ \ \ \ \
+\ \frac{L}{4}\bigg[\frac{(1+\eps)^{-1/2}}{\x-\gt}+
\frac{(1-\eps)^{-1/2}}{\x+\gt}\bigg]\sqrt{a\x^2+b\x+c}-\pi n\,,\nonumber
\end{eqnarray}
where $L=J_1+J_2$   and with  $\epsilon$, $a$, $b$ and $c$ to be determined.
In order to cancel the poles, $a$, $b$ and $c$ must satisfy
$1=\frac{g^2}{2}\,a\,+c, \
b=\frac{\sqrt{2}}{g}\,\epsilon$
while matching the asymptotics gives
\begin{eqnarray}
\pi n=\frac{L\sqrt{a}}{4}\left(\frac{1}{\sqrt{1+\epsilon}}+
\frac{1}{\sqrt{1-\epsilon}}\right) \ , \ \ \ \
\pi (n-2m)=\frac{\sqrt{2}L\sqrt{c}}{4g}\left(\frac{1}{\sqrt
{1+\epsilon}}-\frac{1}{\sqrt{1-\epsilon}}\right)\nonumber
\end{eqnarray}
In the limit $n\to\infty$, one finds
\begin{eqnarray}
&&\epsilon=\frac{\sqrt{\lambda} m}{\sqrt{J_1^2 + m^2\lambda }} \ , \
 \ \ \ \ \ \ \ \qquad\qquad
b=\frac{4\pi m}{\sqrt{J_1^2 + m^2\lambda }}\ , \\
&&a=\frac{1}{2}\,\frac{(4\pi m)^2}{J_1^2 + m^2\lambda +J_1\sqrt{J_1^2 + m^2\lambda }}\ , \ \ \ \ \ \
c=\frac{1}{2}\,\frac{m^2\lambda }{J_1^2 + m^2\lambda +J_1\sqrt{J_1^2 + m^2\lambda }} \ , \nonumber
\end{eqnarray}
where we used the fact that $L/n=J_1/m$ in the limit when  $L$ and $n$ both approach $\infty$.
In this limit the cut shrinks to a point with support at $\x=\x_0$, where
\begin{equation}
\label{y0eq2}
\x_0=\frac{1}{4\pi m}\left(\sqrt{J_1^2 + m^2\lambda }+J_1\right)\,, \ \ \ \ {\rm i.e.} \ \ \ \
y_0=\x_0+\frac{g^2}{2\x_0}=\frac{1}{2\pi m}\sqrt{J_1^2 + m^2\lambda }\,.
\end{equation}
As the cut shrinks to zero length, the density
approaches $\rho(y)=J_1\delta(y-y_0)$ and so $E-J_2$
approaches the same value as in  \rf{ey} (with $k$ in \rf{ey} replaced by $m$
in the notation of the present  section)
\begin{equation}\label{E-J_22}
E-J_2=J_1+2g^2\int dy\,\frac{J_1\delta(y-y_0)}{y\sqrt{y^2-2g^2}+y^2-2g^2}
=\sqrt{J_1^2+ m^2\lambda}\,.
\end{equation}
Note that (\ref{y0eq2}) and (\ref{E-J_22}) are precisely the
 limiting values of,  respectively,  (\ref{y0eq})  and (\ref{E-J_2})
  in the limit $n\to\infty$ if $p=2\pi m$.  In other words,
  this limit of the  circular string
  can be interpreted as a bound state of  $n$ magnons with each magnon
  having $1/n$ of the total energy and momentum.

\bigskip

One can also give a similar interpretation
to the limit of
 pulsating string solutions discussed in \cite{min1,kmmz, kut}.
 The corresponding state is  outside the $SU(2)$ sector on the gauge side
    but is still described by finite gap equations for a string on $R \times S^3$.
 We can write the ansatz for the pulsating
 string solution in terms of the complex coordinates $\XX_1$ and $\XX_2$ as
 (cf. \rf{circ})
\begin{equation}
\XX_1=\sin\theta \ e^{im\sigma}\ , \qquad\qquad \XX_2=\cos\theta\ e^{i\vp}\,,
\ \ \ \ \ \
\theta=\theta(\tau), \ \ \ \ \ \vp= \vp  (\tau) \ .
\end{equation}
This  ansatz corresponds
 to a circular string wrapped $m$ times and with  its  center of mass moving
  along the $\vp$
 direction  with momentum $J$, and which  is pulsating back and forth along $\theta$.  The string equations of motion lead to
\begin{equation}
\dot \vp  =\frac{J}{\sqrt{\lambda}\cos^2\theta}\,,
\end{equation}
which applied to the conformal  constraint gives
\begin{equation}\la{koi}
\kappa^2=\dot\theta^2+m^2\sin^2\theta+\frac{J^2}{\lambda\cos^2\theta}\,.
\end{equation}
If we now assume that $J/\sqrt{\lambda}\gg 1$, and $m\gg 1$ with
$m/J$ fixed, and  further assume that $\theta \ll 1$, the
pulsating becomes harmonic and the constraint equation \rf{koi} is well
approximated by
\begin{equation}
\frac{E^2-J^2}{\lambda}=\dot\theta^2+\left(m^2+\frac{J^2}{\lambda}\right)\theta^2\,.
\end{equation}
Further assuming that $E-J$ is held fixed and
following the analogy    with the standard
harmonic oscillator quantization ($ \epsilon = \hbar \omega N$
where here  $ \omega^2 = m^2+\frac{J^2}{\lambda}$)
 we find that
\begin{equation}\label{E-Jpulse}
E-J\approx \sqrt{\left(\frac{mN}{J}\right)^2\lambda+N^2}\,,
\end{equation}
where $N$ is the oscillator mode number which must satisfy $N\ll
J$ in order that $\theta\ll 1$.

The result (\ref{E-Jpulse}) can also be reproduced from solutions
of the finite gap equation in \cite{kmmz}. In \cite{kmmz} it was shown
that the resolvent arising from the pulsating solution is
\begin{equation}
G(\x)=\frac{1}{2}\,\frac{1}{\x^2-g^2/2}\left(E \x+\sqrt{\left[2\pi m(\x^2-g^2/2)-
J\x\right]^2+(E^2-J^2)\x^2}\,\right)-\pi m\,.
\end{equation}
This  resolvent clearly has four branch points and two cuts.  If we now take
the limit $E,J\to\infty$
  with $E-J$ and $J/m$ finite,
   then the two branch cuts each shrink to a point at
\begin{equation}
\x=\frac{1}{4\pi}\bigg[\frac{J}{m}\pm\sqrt{\left(\frac{J}{m}\right)^2+
\lambda}\bigg]\,,
\ \ \ \ {\rm  i.e.} \ \ \ \ \ \
y=\pm\frac{1}{2\pi}\sqrt{\left(\frac{J}{m}\right)^2+\lambda}=\pm\, y_0\,.
\end{equation}
Hence, the solution has reduced to two zero length condensates which are
 images of each other.  The
densities  along the condensates are opposite to each other so that $J_1=0$.  Each condensate contributes half the oscillator number, so
\begin{equation}
\rho(y)=\frac{N}{2}\Big(\delta(y-y_0)-\delta(y+y_0)\Big)\,.
\end{equation}
The total momentum in (\ref{rhocondy}) must be zero, which means we should choose the branches
$\sqrt{(\pm y_0)^2-2g^2}>0$.  Finally, the third equation in (\ref{rhocondy}) leads to
\begin{eqnarray}
E-J&=&\frac{\lambda }{4\pi}\left(\frac{N}{\left(\frac{J}{m}\right)^2+\frac{J}{m}\sqrt{\left(\frac{J}{m}\right)^2+\lambda}}-\frac{N}{\left(\frac{J}{m}\right)^2-\frac{J}{m}\sqrt{\left(\frac{J}{m}\right)^2+\lambda}}\right)\nonumber\\
&=&N\sqrt{\left(\frac{m}{J}\right)^2+\lambda}\,,
\end{eqnarray}
reproducing (\ref{E-Jpulse}).

We can also work backward and find giant magnon solutions in the pulsating sector.   These solutions would correspond to condensates of equal length and opposite density with total oscillator number $N/2$ on each condensate.  If the density is given by $\pm n$ on each condensate,
then the computation goes through exactly as for the $SU(2)$ case, but with $J_1$ replaced by $N/2$ and $J_2$ by $J$.
The two condensates have momentum $\pm p$, so one finds
\begin{equation}
E-J=2n\sqrt{\left(\frac{N}{2n}\right)^2+\frac{\lambda}{\pi^2}\sin^2\frac{p}{2n}}\,.
\end{equation}
We can reduce this to (\ref{E-Jpulse}) by taking $n\to\infty$ and
identifying $p=mN\pi/J$.

In figure 3 we show the contours for (a) the limit of the folded string and (b) the analogous configuration for a pulsating string.  The distinction between these two cases is that the folded string has both condensates on the same sheet, while the pulsating string has its condensates on different sheets.  The string motion in (b) can be viewed as follows: for half the string, say from $0<\s<\pi$, the configuration is exactly the same as the limit of the folded string, with the string having constant angular velocity along $\varphi_1$.  On the other half of the string everything is the same, except the angular velocity along $\varphi_1$ is in the opposite direction.  Even though the separate halves are rotating in opposite directions in $\varphi_1$, the string is continuous since the two halves are attached where $\cos\theta=0$.   Thus, the string oscillates between a folded configuration and a circular configuration twice every revolution in $\varphi_1$.
\begin{figure}[t]
\centerline{\includegraphics[scale=.8]{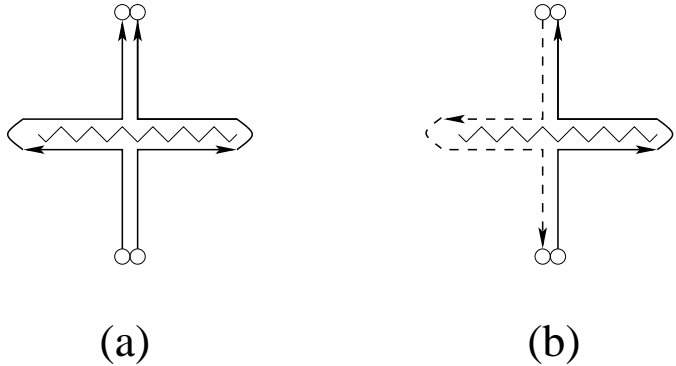}}\caption{Condensates
for strings made up of two giant magnons.
 (a)  is the limit of the folded string  and
 (b) is a  pulsating string.
  The arrows represent the sign of the density while the dashed line in (b) indicates that the condensate is on the lower sheet.} \nonumber
\end{figure}
\bigskip

In Appendices  B  and  C we shall also discuss similar solutions in the
$SL(2)$ sector.

\bigskip
\bigskip
While this  paper was in preparation we learned of
an interesting  forthcoming paper \ci{afz}  that discusses the
  finite $J$ generalization
  of the giant magnon solutions of \ci{hm,dor1}.


\bigskip
\bigskip

\section*{Acknowledgments }

We are  grateful to  G. Arutyunov, S. Frolov,
 I. Klebanov, M. Kruczenski,  J. Russo  and M. Zamaklar
 for useful discussions.
  The  work of J.A.M. was supported in part by the Swedish Research Council.
 The work of A.T. and A.A.T.
  was supported  in part by the DOE grant DE-FG02-91ER40690.  A.A.T.
acknowledges also the support of
 INTAS  03-51-6346 and  EC MRTN-CT-2004-005104  grants
  and the RS Wolfson award.

\renewcommand{\theequation}{A.\arabic{equation}}
\renewcommand{\thesection}{A}
 \setcounter{equation}{0}
\setcounter{section}{1} \setcounter{subsection}{0}

 \section*{Appendix A:   Fluctuation Lagrangian near the folded string solution  }

\subsection{Bosonic fluctuations}

Our starting point will be the general form of the 2-spin folded string
solution  discussed in section 2.2. We shall  consider the conformal gauge.

Since the string is not stretched in the spatial $AdS_5$ directions
(with the metric $
ds^2=-\frac{(1+\frac{1}{4}\zeta^2)^2}{(1-\frac{1}{4}\zeta^2)^2}dt^2+
\frac{d\zeta_{k}d\zeta_{k}}{(1-\frac{1}{4}\zeta^2)^2}$)
their fluctuations
$
t=\kappa\tau+\tilde{t}, \  \zeta_{k}=0+ \tilde{\zeta_{k}}, \ k=1,2,3,4
$
are governed by
\begin{equation}
L=-\frac{1}{2}\big[-(\partial_{a}\tilde{t})^2
+(\partial_{a}\tilde{\zeta_{k}})^2+\kappa^2\tilde{\zeta}_{k}^2\big]\ ,
\label{fluctads}
\end{equation} i.e. we get one
 massless fluctuation and $4$
 massive ones  with the characteristic frequencies
$\omega =\pm \sqrt{n^2+\kappa^2}$.

To consider the $S^{5}$ fluctuations we shall
 follow \ci{art,ptt} and use complex  embedding coordinates
 in terms of which the $S^5$   Lagrangian is
\begin{equation}
L=-\frac{1}{2}\partial_{a}\X_i \partial^{a}\X^*_i+\frac{1}{2}\Lambda
(\X_i \X^*_i -1)    \ ,
\end{equation}
 and  the classical solution is
\begin{equation}
\X_1= \cos \theta(\sigma) \ e^{i w_1  \tau}, \quad \X_2= \sin \theta(\sigma) \ e^{i w_2
\tau}, \quad  \X_3=0 \label{folded3} \ ,
\end{equation}
so that  the classical value of the Lagrange
multiplier is
\begin{equation}\la{lya}
\Lambda = \partial_{a}\X_i \partial^{a}\X^*_i  =
  -2  (\kappa^2-w^2_1) \frac{\sin^2 \theta}{\sin^2 \theta_*}-2 w^2_1 +\kappa^2\ .
\end{equation}
Introducing the fluctuations $ \X_i \to \X_i + \tilde{\X}_i$  one gets
 \bea
 \td {L}=-\fr12\;\pa_a\tilde{\X}_i\pa^a\tilde{\X}_i^*
 +\fr12 {\Lambda}
  \tilde{\X}_i\tilde{\X}_i^*\ , \ \ \ \ \ \ \ \ \ \ \
 \sum_{i=1}^{3}(\X_i \tilde{\X}_i^*+\X_i^* \tilde{\X}_i)=0\ .
 \label{str}
 \eea
  $\tilde\X_3$ has no classical background and thus decouples, i.e. its
 equation of motion is
 \begin{equation}
 \partial_{0}^2\tilde{\X}_3-\partial_{1}^2\tilde{\X}_3-\Lambda \tilde{\X}_3=0\ ,
 \label{x3}
 \end{equation}
 where $\L=\L(\s)$.

The remaining 3 independent  fluctuations are coupled. Let us define
 \be
 \tilde{\X}_1=e^{iw_1 \tau}(g_1+if_1)\ , \ \  \
  \tilde{\X}_2= e^{i w_2 \t}(g_2+if_2)\ ,
 \ee
where the  constraint in  \rf{str}  implies
\begin{equation}
g_1 \cos \theta+g_2 \sin \theta=0 \label{const} \ .
\end{equation}
Then
\begin{eqnarray}
\td L&=&\frac{1}{2} \bigg[
\dot{g}_1^2+\dot{g}_2^2+\dot{f}_1^2+\dot{f}_2^2-g_1'^2
-g_2'^2-f_1'^2-f_2'^2+
w^2_1 (f_1^2+g_1^2)+w_2^2 (g_2^2+f_2^2)\nonumber\\
&-&4 w_1 f_1 \dot{g}_1-4 \w_2  f_2 \dot{g}_2 +  \Lambda
(f_1^2+f_2^2+g_1^2+g_2^2)  \bigg] \label{fluct1}
\end{eqnarray}
We can simplify this by introducing
\begin{equation}
\xi=g_1 \cos \theta+g_2 \sin \theta,\ \ \  \quad \eta=-g_1 \sin
\theta+g_2 \cos \theta \label{rot1}\ ,
\end{equation}
and   (\ref{const})  implies that  $\eta_{1}=0$.
The fluctuation Lagrangian
for $f_1,f_2,\eta$ then
 becomes
\begin{eqnarray}
\td L&=&\frac{1}{2}\bigg[\dot{f}_1^2+\dot{f}_2^2-f_1'^2-f_2'^2 +
\dot{\eta}^2-\eta'^2- M^2_\eta  \eta^2 - M^2_{1} f_1^2-
M^2_{2}f_2^2
\nonumber\\
&+& 4 (w_1 \sin \theta \ f_1-
\frac{\sqrt{\k^2 -w^2_1 \cos^2 \theta_*}}{\sin \theta_*} \cos \theta\
f_2  ) \dot{\eta}
 \bigg] \ ,  \label{fluct2}
\end{eqnarray}
where
\begin{equation}
M^2_\eta =    - (\kappa^2 - w^2_1 )       \frac{\cos 2\theta}{\sin^2 \theta_*},
\end{equation}
\begin{equation}
M^2_{1} = - (\kappa^2 - w^2_1 )\bigg(1-2\frac{\sin^2 \theta}{\sin^2
\theta_*}\bigg) \ , \ \ \ \ \ \ \ M^2_{2} =    -   (\kappa^2 - w^2_1 )
\bigg(1+\frac{\cos 2\theta}{\sin^2 \theta_*}\bigg) \ ,
\end{equation}
and   we used the explicit form of $w_2$ from  (\ref{rg}).

\def \X {{\cal X}}

\subsection{Fermionic fluctuations}

 The  quadratic part of the \adss
superstring  Lagrangian evaluated on a bosonic solution has  a
simple form (see \ci{mt,ft1,ft2,ft3} for details)
 \bea
 L_F=i\left(\eta^{ab}\delta^{IJ}-\e^{ab}s^{IJ}\right)\;
 \bar{\vartheta}^I\rho_aD_b\,\vartheta^J\;\;,\;\;\ \ \ \rho_a\equiv \G_A
 e_a^A\ , \;\;\;\; e_a^A\equiv E_\m^A(\X)\pa_a \X^\m\ ,
 \eea
where $I,J=1,2,\;s^{IJ}=\mbox{diag}(1,-1),$ \ $ \rho_a$ are
projections of the ten-dimensional Dirac matrices and $\X^\m$ are
 the coordinates of the $AdS_5$  space for $\m=0,1,2,3,4$ and
 the coordinates of $S^5$
for $\m=5,6,7,8,9$. The covariant derivative is given by
 \bea
 && D_a\vartheta^I=\left(\delta^{IJ}{\rm D}_a-\fr{i}2\e^{IJ}\G_*\rho_a\right)\vartheta^J
 \;,\ \ \ \ \quad  \G_*\equiv i\G_{01234}\;,\;\ \G_*^2=1 \ ,
\eea
where ${\rm D}_a=\pa_a+\fr14\w_a^{AB}\G_{AB}, \ \  \w_a^{AB}\equiv
 \pa_a  \X^\mu  \w_\mu^{AB}$.
Fixing the $\k$-symmetry by the same condition as in \ci{ft3} \
 $\vartheta^1=\vartheta^2=\vartheta $ \
one gets
 \bea \la{fe}
  L_F=-2i\bar{\vartheta}D_F\vartheta\;,\quad\quad  D_F=-\rho^a {\rm
  D}_a-\fr{i}2\e^{ab}\rho_a\G_*\rho_b\ .
 \eea
Labelling the coordinates as follows:
 \bea
\m: && 0\;\;1\;\;2\;\;3\;\;\;4\;\;\;\;5\;\;\;6\;\;\;\;7\;\;\;8\;\;\;\;9\nonumber\\
\X^\m: &&
t\;\;\rho\;\;\psi\;\;\phi_1\;\;\phi_2\;\;\g\;\;\varphi_3\;\;\theta\;\;
 \varphi_1\;\;\varphi_2
 \eea
we find that in  the case of the folded solution
that the
non-trivial components of the Lorentz connection $\w_a^{AB}$ are
\begin{equation}
 \w_0^{87}=- w_1 \sin \theta,\ \ \ \ \  \quad \w_0^{97}= w_2  \cos
 \theta\ .
 \end{equation}
For $\rho_a$ we find
\begin{equation}
\rho_0=\kappa \Gamma_0+  w_1  \cos \theta \Gamma_8+ w_2
\sin \theta \Gamma_9, \quad  \ \ \ \ \    \rho_1= \Gamma_7 \theta'
\end{equation}
The operator $D_F$ becomes
\begin{eqnarray}
D_F&=& (\kappa \Gamma_0+ w_1 \cos\theta \Gamma_8+ w_2 \sin\theta
\Gamma_9)
\partial_{0}- \Gamma_7 \theta'
\partial_{1}\nonumber\\
&-& \frac{1}{2}(\k \Gamma_0+w_1 \cos \theta \Gamma_8+ w_2 \sin
\theta \Gamma_9)(w_1 \sin \theta \Gamma_{87}-w_2 \cos \theta
\Gamma_{97})\nonumber\\
&+ & \theta'(w_1 \cos \theta \Gamma_8 +w_2 \sin \theta
\Gamma_9)\Gamma_{07}\Gamma_{1234} \label{Ffluct1}
\end{eqnarray}
In section 3.1  we shall  consider the special limit of this operator when
$\theta_*=\frac{\pi}{2}$ and $\k \to \infty$.

\bigskip

\subsection{Some details}

In the main text we   also use   the  general expression for the 1-loop correction to the energy
in terms of the  bosonic and fermionic characteristic frequencies  \ci{ptt}
 \begin{eqnarray}
E_1=\fr1{\k}E_{2{\rm d}}&=&\fr{1}{2\k}\bigg[ \sum_{p=1}^{8}\left(
     \hat w_{p,0}^B -\hat \w_{p,0}^{F}
    \right)
   +   \sum_{n=1}^\infty\;\sum_{I=1}^{16} \bigg(
    \hat\w_{I,n}^B- \hat\w_{I,n}^{F}\bigg) \bigg] \ ,
 \label{e1}
 \end{eqnarray}
 \be
\hat  \w_{p,0}
 =   sign(C_{p}^{B}) \w_{p,0}\ , \ \ \ \ \ \ \ \
\hat \w_{I,n} =  sign(C_{I,B}^{(n)})  \w_{I,n} \ , \ee
 \begin{equation}
C_{p}^{B}=\frac{1}{2m_{11}(\w_{p,0})\w_{p,0} \prod_{q\neq
p}(\w_{p,0}^{2}-\w_{q,0}^{2})},\quad
C_{I,B}^{(n)}=\frac{1}{m_{11}(\w_{I,n})\prod _{J\neq
I}(\w_{I,n}-\w_{J,n})}\ ,  \label{signs}
\end{equation}
where $m_{11}$ is a minor of $F$, i.e. the determinant of the
matrix obtained from $F$ by removing the first row and first
column, with $F$ being the matrix entering the equation $\det F=0$
for the
characteristic frequencies. This matrix satisfies the condition
$F^{T}( \om_{I,n},n)=F(- \om_{I,n},-n)$ (see \cite{ptt} for
details).

\bigskip \bigskip

Let us also explain   how one arrives at eq.\rf{cfreq}  of section 3.1,
and, in particular,  why one  can  indeed
ignore the  contribution of the $x=0$ point. From (\ref{fluct4}) we get
\bea
&&\ddot{f}_1-f_1''-(w^2-1) f_1+ 2\epsilon(x)  w \dot\eta =0\ ,\\
&&\ddot{\eta} -\eta ''- (w^2-1)\eta - 2\epsilon(x)  w \dot{f}_1=0 \ ,
\eea
and looking  for  solutions $
f_1\sim A(x)e^{i\omega t},\ \
\eta\sim B(x)e^{i \omega t}$
we get
\begin{equation}
A'' + (\omega^2 + w^2-1)A-  2i \omega w  \epsilon(x)   B=0
\ , \ \ \ \ \ \ \
B'' + (\omega^2 +  w^2-1)B+  2i \omega w \epsilon(x)   A=0
\end{equation}
Combining these  two equations we get a 4-th  order differential
equation for $A$,
which
(after using that $\delta (x)   \epsilon(x)=0$)
 becomes
\begin{eqnarray}
&&\epsilon^2(x) [A''''+\omega^2 A''+(w^2-1)A'']-4 \epsilon^4(x) \omega^2
w^2 A+ \delta(x)[\omega^2
A+A''+(w^2-1)A]\nonumber\\
&&+\ \epsilon^2(x)(\omega^2+w^2-1)[\omega^2 A+A''+(w^2-1)A]=0
\label{4order}
\end{eqnarray}
We can solve this equation for $x<0$ and $x>0$ with the ansatz
$A\sim e^{ip x}$ and obtain the characteristic frequencies
(\ref{cfreq}). Notice that the equation (\ref{4order}) contains a
delta-function term  which signals a discontinuity at the origin.
Integrating (\ref{4order}) near $x=0$ and taking the interval
of integration to zero we find that the  only non-vanishing term is
\begin{equation}
\omega^2 A(0)+A''(0)+(w^2-1)A(0)=0\ .  \label{condition}
\end{equation}
One can see that one cannot have the  solution $A\sim e^{ip x}$
valid at the origin since the frequencies (\ref{cfreq}) do not
satisfy equation (\ref{condition}) unless $w=0$.
To satisfy
(\ref{condition}) also for $w\neq 0$ we need to have $A(0)=0$.
This shows that $A(x)$ is discontinuous at origin. Therefore,
 one can just ignore the $x=0$ point and  thus  obtain
  (\ref{cfreq}).

\renewcommand{\theequation}{B.\arabic{equation}}
 \setcounter{equation}{0}
\setcounter{section}{1} \setcounter{subsection}{0}
\renewcommand{\thesection}{B}

 \section*{Appendix B:   Large $J$  limit of circular $(S,J)$  solution in
  the  $SL(2)$  sector     }

It is straightforward  to perform the analog of the analysis
of sections 2.3   and 3.2  and  consider the $J \gg S$ limit
of the  circular 2-spin solution in the $SL(2)$ sector \ci{art,ptt}.
One finds again the square root formula for the classical energy
similar to \rf{er} and also that 1-loop correction to it vanishes.

\renewcommand{\theequation}{B.\arabic{equation}}
\subsection{ Limit of classical solution }

Let us start  with a review of
the solution \cite{art,ptt} describing circular string which is
rotating both in $AdS_5$ and in $S^5$. In terms of  complex combination of embedding
coordinates one has
 \bea
 \Y_{0}=r_{0} \; e^{i\kappa \tau}\
 , \quad  \quad    \Y_{1}=r_{1} \;e^{i\varpi \tau+i m\sigma}\ ,\quad
  \quad \XX_1=e^{iw\tau+ik\sigma}\ , \quad \Y_2, \XX_2,\XX_3=0
  \label{sol}
 \eea
 \be \la{rad}
 r_0 \equiv \cosh \rho_{0}\ , \quad\quad \ \ \ \
 r_1\equiv \sinh \rho_{0}\ , \ \ \ \ \ \ \ \ \ \ \ \ \
  r_0^2 - r_1^2 =1 \ .
 \ee
Here $\rho_{0}$ is a  constant radius of the circular string
in $AdS_5$,\  $k$
and $m$ are the winding numbers, and  $\varpi$ and $w$ are rotation frequencies of the
string.
From equations of motion   we have
 \bea
 && \varpi^{2}=\kappa^2+   m^2, \qquad w^2=\nu^2 + k^2 , \qquad
 \nu^2= -\Lambda, \qquad \k^2=\tilde{\Lambda} \ ,  \label{pararel}
\eea
where  $\Lambda$  and $ \tilde{\Lambda}$ are the Lagrange multipliers for the embedding
coordinates. The energy and the two  non-zero spins are
\begin{equation}
E= \sql \E = \sql  r^2_0\kappa\ , \qquad  S= \sql \S=\sql  r^2_1 \varpi  \ , \qquad
J= \sql \J=\sql  w\ ,
\end{equation}
and  the conformal gauge constraints imply
\be 2\kappa \mathcal{E}-\kappa^2=2\sqrt{\kappa^2 + m^2 }\mathcal{S}
+\mathcal{J}^2+k^2\ ,
 \label{cgc}\ee
\be \la{kik}
 m \mathcal{S}+k \mathcal{J}=0\ ,
\ee
while  \rf{rad} gives  also \be
\frac{\mathcal{E}}{\kappa}-\frac{\mathcal{S}}
{\sqrt{m^2+\kappa^2}}=1 \  . \la{rell}\ee
Eliminating $\k$  from \rf{cgc} and \rf{rell} one  finds $\E=\E(\S,\J,m)$.

Let us  now  consider the special limit when $\mathcal{J}\to
\infty$  with $\S$  and $k$ being fixed and negative (this implies
$m\gg 1 $). Then $\E$  is also  divergent but $\E-\J$ is  finite.
We get
\begin{equation}
\kappa=  \frac{\J}{|k|}  +\frac{k^2}{\sqrt{k^2+\mathcal{S}^2}}+O({
1 \ov \J} ) \ ,
\end{equation}
\begin{equation}
r_0=1+\frac{\mathcal{S}^2 }{2\sqrt{k^2+\mathcal{S}^2}}\frac{1}{\J
}+...\ , \ \ \ \ \ \ \ \quad r_1=\frac{\mathcal{S}
}{(k^2+\mathcal{S}^2)^{1/4} }\frac{1}{\sqrt{\J}}+...,
\end{equation}
\begin{equation}
\varpi=\frac{\J}{\S}\sqrt{k^2+\mathcal{S}^2}+
\frac{\mathcal{S}k^2}{k^2+\mathcal{S}^2}+...
\end{equation}
and  finally in the limit of $\mathcal{J}\rightarrow \infty$
\begin{equation}
E-J=\sqrt{S^2+k^2\lambda} \ .
\end{equation}

\subsection{Vanishing of $1$-loop correction to  classical energy }

Let us set $k=-1$ for simplicity. For generic $\J$ and $\S$ the
 bosonic and fermionic fluctuation
 frequencies were obtained in \cite{ptt}. There are
$4$ real free massive fields with mass $\nu$, for which in the
limit (and after the rescaling of the coordinates $t=\k \tau, \ x=
\k \s$)  we get $ \omega=\pm \sqrt{p^2+ 1 }.  $ There are also two
free massive modes with mass $\kappa$, which in the limit has the
same frequencies. The remaining coupled fluctuation Lagrangian in
the large $m$-limit reads (cf. \rf{fluct3})
\begin{equation}
\bar L=\frac{1}{2}\bigg(\dot{f}_1^2-f_1^{'2}+\dot{F}_0^2-F_0^{'2}+\dot{F}_1^2-F_1^{'2}+
\dot{G}_1^2-G_1^{'2}\bigg)-2 \sqrt{1+\mathcal{S}^{-2}}F_1 \dot{G}_1+2
F_1 G_1' \ ,
\end{equation}
where $F_0$, $F_1$ and $G_1$ are fluctuations in $AdS_5$ directions. The
non-trivial characteristic frequencies are found to be similar to
the ones in the $SU(2)$ case  (cf. \rf{w13})
\begin{equation}
\omega_{1,2}=\sqrt{1+\b^2}\pm
\sqrt{(p+ \b)^2 + 1 } \ ,\ \ \ \ \ \ \ \
\omega_{3,4}=-\sqrt{1+\b^2}\pm
\sqrt{(p- \b)^2 + 1 } \ ,
\label{l2}
\end{equation}
$$ \b \equiv \S^{-1}  \ . $$
The fermionic fluctuation  Lagrangian has the following general form
\cite{ptt}
\begin{equation}
L= 2 i \bar{\vt}D_F \vt
\ , \ \ \ \ \ \ \ \ \ \ \
 {D_F}
 =    \G_0 \partial_{0}-\G_3 \partial_{1} \ \pm  i a  \G_{1}
 +c \G_{016}  +   d \G_{136} \ ,
 \label{df}
 \ee
where \bea
 a=\fr{\sqrt{2}m   \k r_0r_1 }{\sqrt{\k^2-\n^2}} \; , \;\;\;\;\;
c=\fr{\k k}{\varpi}\fr{\varpi^2-w^2}{\k^2-\n^2}\;,\;\;\;\;\;\;\;
  \;\; d= \fr{k m\k r_0^2}{{\k^2-\n^2}}\ .
 \eea
Expanding in large $\J$ and rescaling the
coordinates we  obtain for $k=-1$
\begin{equation}
{D_F}
 =    \G_0  \partial_{t}-\G_3  \partial_{x}\pm i
   \G_1-\frac{1}{2} \b \G_{016}  -  \sqrt{1+\b^2}
 \G_{136} \ .
\end{equation}
The resulting  fermionic
characteristic frequencies are
\begin{equation}
\omega=\pm \sqrt{1+\b^2 }\pm
\sqrt{\big(p\pm \frac{1}{2}\b \big)^2  + 1 } \ .  \label{fsl2}
\end{equation}
Proceeding as in the $SU(2)$ sector in section 3.2
 to compute the 1-loop
correction to the energy, we again find
 using (\ref{signs})   that in both the bosonic and fermionic
cases the $p$-independent square roots in (\ref{l2}) and (\ref{fsl2}) do not contribute
to $E_1$.
As a  result, we get
the same integral  \rf{op}  as in the $SU(2)$  case
\begin{eqnarray}
E_1&=&\frac{1}{2}\int_{-\infty}^{\infty}dp\
\bigg[6\sqrt{p^2+1}+\sqrt{(p+\b )^2+1}+
\sqrt{(p-\b )^2+1}\nonumber\\
&-&4 \sqrt{\big(p+\frac{1}{2}\b\big)^2+1}-4
\sqrt{\big(p-\frac{1}{2}\b\big)^2+1}\bigg]=0  \ .
\label{1loopsl2}
\end{eqnarray}

\renewcommand{\theequation}{C.\arabic{equation}}
 \setcounter{equation}{0}
\setcounter{section}{1} \setcounter{subsection}{0}
\renewcommand{\thesection}{C}

 \section*{Appendix C:   Giant magnons in
  the  $SL(2)$  sector     }

In this Appendix we shall
consider ``giant magnons'' in the $SL(2)$ sector, i.e. the analogs  of the solutions
of  \ci{hm} and
of section 2.1 that have spins in both $AdS_5$ and $S^5$.
These ``magnons'' turn   out to stretch
to the boundary of $AdS_5$ and,  strictly speaking,  have
 not only infinite energy, but also infinite $E-J$.
 However,
this infinity, unlike the usual infinity for $E$ or
$J$ is associated with the boundary, and as such can be
removed with a local counterterm.  The  final result is finite.

The setup is similar to the $SU(2)$ case in section 2.1.   The relevant metric is
that of $AdS_3 \times S^1$ part of \adss
\begin{equation}
ds^2=-\cosh^2\rho\ dt^2+d\rho^2+\sinh^2\rho\  d\chi^2+d\phi^2\,,
\end{equation}
and we make the ansatz
 \begin{eqnarray}
 t&=&\tau\ , \qquad\qquad \phi=t+\varphi(\s) \, \nonumber\\
\rho&=&\rho(\s)\ ,  \qquad \chi=w\,(t -\psi(\s))\,.
 \end{eqnarray}
 We then find that $\DD$ in the action \rf{acd}  is given by
 \begin{eqnarray}
 \DD&=&(\cosh^2\rho -1-w^2\sinh^2\rho)((\p_\s\varphi))^2+w^2\sinh^2\rho\ (\p_\s\psi)^2+(\p_\s\rho)^2\nonumber\\
 &&\qquad\qquad\qquad+(\p_\s\varphi-w^2\sinh^2\rho\ \p_\s\psi)^2\nonumber\\
&=&\cosh^2\rho\,(\p_\s\varphi)^2+w^2\sinh^2\rho\,\cosh^2\rho\,(\p_\s\psi)^2+(1-w^2)\,\sinh^2\rho\,(\p_\s\rho)^2\nonumber\\
&&\qquad\qquad\qquad-w^2\sinh^2\rho\,(\p_\s\varphi+\p_\s\psi)^2 \ .
 \end{eqnarray}
The resulting equations of motion have  the special solution for  $\psi$
 \begin{equation}
\p_\s\psi=\frac{1}{\sinh^2\rho} \p_\s\varphi\,.
  \end{equation}
  Substituting it back into the action we have the same expression
  as in (\ref{redact}), except that now
  \begin{equation}
r=\cosh\rho =\frac{\sin\varphi_0}{\sin\varphi}\ , \qquad\qquad -\varphi_0<\varphi<\varphi_0\,.
  \end{equation}
The difference $E-J$ and the spin $S$ are then given by
\begin{eqnarray}\label{sl2ejs}
E-J&=&\frac{\sqrt{\lambda}}{2\pi\sqrt{1-w^2}}\int_{-\varphi_0}^{\varphi_0}d\varphi \frac{\sin\varphi_0}{\sin^2\varphi}\nonumber\\
S&=&w(E-J)\,.
\end{eqnarray}
Strictly speaking, the quanties in (\ref{sl2ejs}) are infinite
because of the singularity at $\varphi=0$.  This corresponds
to $\rho=\infty$ which is at the boundary of $AdS_5$.  Hence,  this divergence is in the UV and differs from the individual divergences of $E$ and $J$ which are in the IR. Accordingly, the divergence can be cancelled with a counterterm.  This is accomplished by deforming the contour slightly away from $\varphi=0$, giving the regulated answers
\begin{eqnarray}
(E-J)_{\rm reg}=-\frac{\sqrt{\lambda}\cos\varphi_0}{\pi\sqrt{1-w^2}}\ , \ \ \ \ \ \ \ \ \ \
S_{\rm reg}=-\frac{w\sqrt{\lambda}\cos\varphi_0}{\pi\sqrt{1-w^2}}\,,
\end{eqnarray}
where the subscript (reg) refers to the regulated quantities.
We can then write
\begin{equation}\label{E-JSL2}
(E-J)_{\rm reg}=-\sqrt{|S_{\rm reg}|^2+\frac{\lambda}{\pi^2}\sin^2\frac{p}{2}}\,.
\end{equation}

One can also derive  this result using the finite gap analysis.
 We first remark that an $SL(2)$ spin chain, strictly speaking,
 cannot have Bethe strings of finite size.  For example, the
  Bethe equations for the one loop anomalous dimension in the $SL(2)$ sector are
\begin{equation}\label{SL2bethe}
\left(\frac{y_j-i/2}{y_j+i/2}\right)^J=\prod_{k\ne j}^S\frac{y_j-y_k+i}{y_j-y_k-i}\,.
\end{equation}
In the limit $J\to\infty$, the left hand side is zero if ${\rm Im}\, y_j>0$.
  This means that the right hand side must also be zero, which can be
   accomplished only if there is also a root at $y_j+i$.  But then replacing
    by $y_j$  by $y_j+i$ in the l.h.s. of (\ref{SL2bethe}) we again end up with
    a zero, which means that there is a root at $y_j+2i$,  and the
    argument continues {\it ad infinitum}.  Hence, there are an infinite
    number of roots in the string and so $S$ is infinite.

When taking the continuum limit, the Bethe equations turn
into integral equations and the Bethe strings become condensates.
 In the finite gap equations
 this translates into  condensates  of infinite extent.
   Furthermore, in order for the energies to be real,
    every infinite condensate must be paired with its complex conjugate.
The finite gap equations for the $SL(2)$ sector are very
similar to the $SU(2)$ equations \cite{KZ}, and,   in particular,
the equations in (\ref{rhocondy}) are the same with $J_1$
 and $J_2$ replaced by $S$ and $J$.  Hence, we find that for an infinite condensate and its conjugate
\begin{equation}
S=\int_{-i\infty +y_0}^{+i\infty+y_0}dy\ \rho(y)-\int_{-i|S|/2 +y_0}^{+i|S|/2+y_0}dy\ \rho(y)\,.
\end{equation}
The first integral is infinite if $\rho=-i$ along the path.  However, if we deform the contour slightly the integral will be zero, since $\rho$ only has  a double pole at infinity.  Hence we find
$S_{\rm reg}=-|S_{\rm reg}|$.  Likewise,
\begin{equation}\label{SL2E-J}
(E-J)_{\rm reg}=S_{\rm reg}-2g^2\int_{-i|S|/2+y_0}^{+i|S|/2+y_0} dy\,\frac{\rho(y)}{y\sqrt{y^2-2g^2}+y^2-2g^2}\,.
\end{equation}
We solve for $y_0$ the same way as in section 4.3 and then (\ref{SL2E-J}) immediately gives
(\ref{E-JSL2}).

The same result can be derived for the $SL(2)$ sector
 from the discrete asymptotic  BDS-type  Bethe equations
 in \ci{Staudacher:2004tk,Beisert:2005fw}.  The arguments
 work in almost the same way as for the $SU(2)$ sector as
 discussed in section 4.   In this case the Bethe equations become
\begin{equation}\label{SL2beth}
\left(\frac{\x_j^-}{\x_j^+}\right)^{J}=\prod_{k\ne j}^{S}
\frac{y_j-y_k+i}{y_j-y_k-i}\left(\frac{1-
{g^2\ov 2\x_j^-\x_k^+} }{1-{g^2\ov 2\x_j^+\x_k^-}}\right)^2\,,
\end{equation}
where $\x_j^\pm=\x(y_j\pm i/2)$.
Hence, as in the one-loop case if ${\rm Im}\,y_j>0$,
then there must be a root at $y_j+i$.  Hence, the Bethe
string goes on forever in the imaginary direction. In order
to have real solutions, we require that there also be the
complex conjugate of this Bethe string.  In any case, one now finds that
\begin{eqnarray}
(E-J)_{\rm reg}-S_{\rm reg}&=&ig^2\sum_{j=1}^{\infty}
\left(\frac{1}{\x(y_0+i|S_{\rm reg}|/2+ij)}-\frac{1}{\x(y_0+i|S_{\rm reg}|/2+ij-i)}\right)\nonumber\\
&&+ig^2\sum_{j=1}^{\infty}\left(\frac{1}{\x(y_0-i|S_{\rm reg}|/2-ij+i)}-\frac{1}{\x(y_0-i|S_{\rm reg}|/2+ij)}\right)\nonumber\\
&=&-ig^2\left(\frac{1}{\x(y_0+i|S_{\rm reg}|/2)}-\frac{1}{\x(y_0-
i|S_{\rm reg}|/2)}\right)\,.
\end{eqnarray}
This then leads to (\ref{E-JSL2}).

\bigskip

The  negative sign in front of the square root in (\ref{E-JSL2})
may seem puzzling,  so
let us try to give a possible
interpretation of this configuration on the gauge side.
The divergence of  $S$ and $E-J$ is due to the string going out
 to the boundary of $AdS_5$.  This suggests that
 we have inserted a localized adjoint gauge source,
  in other words, a Wilson line in the adjoint
  representation along a particular trajectory of the gauge theory.
   The infinite value for $E-J$ can then be interpreted
   as the infinite contribution coming from a source
   of infinite mass, as was the case for the quark-antiquark
   configuration in \cite{Rey:1998ik,Maldacena:1998im}.
    Likewise, if the source is moving along the boundary,
    it will have infinite angular momentum if it has infinite mass.
      The regularization then corresponds to subtracting off this
      infinite energy and angular momentum and the resulting finite
      $E-J$ and $S$ are the contributions of the operators in the
      presence of these sources.  If one thinks of the boundary
       theory as being defined on $R\times S^3$, then the allowed states
        must be color singlets on $S^3$.  Hence, if an adjoint
    source is inserted somewhere on the $S^3$, this must
     bind onto states such that the net color is zero.\footnote{Let us
     note that in the  Poincare coordinates in $AdS_5$
     with the metric $ds^2=
    \frac{R^2}{z^2} (-d{\rt}^2+dr^2+r^2 d\theta^2+dz^2)$, the
    above solution   has the
    form:
$$
z=\frac{R\sin\varphi}{\sin\varphi_0\cos t}, \quad \rt=R\tan t,
\quad r^2=(R^2+\rt^2) (1-\frac{\sin^2\varphi}{\sin^2\varphi_0}),
\quad \theta = w \arccos (\frac{R}{\sqrt{R^2+\rt^2}}) \ .
$$
The boundary is at $z=0$ which occurs at $\varphi=0$. Here $t$
is the global time and $\rt$ refers to the Poincare patch
time.  The
trajectory at the boundary has the source coming in from infinity
and reaching a minimum distance $R$ at $\rt=0$.  In the meantime its
angle changes between $-\frac{w\pi}{2}$ and $+\frac{w\pi}{2}$ (as
$w$ approaches $1$ the trajectory approaches a lightlike straight
line).}
     With the background color source, we see no
     violation of the usual supersymmetry arguments
      that normally enforce $E\ge J$.

Note that the circular  $SL(2)$ solution discussed
 in Appendix B is  {\it not} made up of magnons of this type.
Instead,  the  circular solution has a single
cut  shrinking to zero size along the real axis, which
 contrasts with the  $SU(2)$ case where it is
a cut along the imaginary direction that is shrinking.
  But the bound
magnons correspond to roots extended along the
 imaginary direction, and so,  unlike the $SU(2)$ case,
 it is not possible to see the $SL(2)$ circular
 solution emerging as a limiting case  of bound magnons.

\vfill


\end{document}